\documentclass[aps,twocolumn,pra,tightenlines,floatfix,showpacs]{revtex4-1}

\usepackage[dvips]{graphicx}
\usepackage[english]{babel}
\usepackage{amsmath}
\usepackage{amssymb}
\usepackage{slashed}
\allowdisplaybreaks

\begin{document}
\title{Density and Spin Linear Response of Atomic Fermi Superfluids with Population Imbalance in BCS-BEC Crossover}
\author{Hao Guo$^{1}$, Yang Li$^{1}$, Yan He$^2$, and Chih-Chun Chien$^3$}
\affiliation{$^1$Department of Physics, Southeast University, Nanjing 211189, China}
\affiliation{$^2$James Franck Institute and Department of Physics,
University of Chicago, Chicago, Illinois 60637, USA}
\affiliation{$^3$Theoretical Division, Los Alamos National Laboratory, Los Alamos, NM, 87545, USA}

\begin{abstract}
We present a theoretical study of the density and spin (representing the two components) linear response of Fermi superfluids with tunable attractive interactions and population imbalance. In both linear response theories, we find that the fluctuations of the order parameter must be treated on equal footing with the gauge transformations associated with the symmetries of the Hamiltonian so that important constraints including various sum rules can be satisfied. Both theories can be applied to the whole BCS-Bose-Einstein condensation crossover. The spin linear responses are qualitatively different with and without population imbalance because collective-mode effects from the fluctuations of the order parameter survive in the presence of population imbalance, even though the associated symmetry is not broken by the order parameter. Since a polarized superfluid becomes unstable at low temperatures in the weak and intermediate coupling regimes, we found that the density and spin susceptibilities diverge as the system approaches the unstable regime, but the emergence of phase separation preempts the divergence.
\end{abstract}

\date{\today}
\pacs{03.75.Ss,74.20.Fg,67.85.-d}

\maketitle

\section{Introduction}\label{Sec.1}
Ultracold Fermi gases provide a clean testbed for many-body theories and act as simulators of complex many-particle systems \cite{StringariRMP08,ZwergerRMP08,Pethickbook}. As the attractive interaction is tuned by varying an external magnetic field, the ground state of a two-component Fermi gas continuously evolves from a BCS superfluid to a Bose-Einstein condensate (BEC) of dimers, a phenomenon called BCS-BEC crossover \cite{Pethickbook,Uedabook}. There has been a broad literature on the linear response of two-component Fermi gases with equal population of the two components \cite{HaussmannPRL12,StrinatiPRL12,HaoPRL10,ValePRL12,OurJLTP13,HHEPL10,HLPRA10,HLPRA12}.  When population imbalance between the two components, usually identified as the two spins \cite{spin_note}, is introduced, interesting phases and phase separation emerge and this has been a subject of intense studies \cite{ZSSK06,Rice1,ChienPRL,SheehyRoPP}.

Although the linear response theory for Fermi superfluids without population imbalance is well developed (see Refs.~\cite{OurJLTP13,OurIJMPB} and references therein), there are few studies on the linear response of population-imbalanced systems and past work relies on techniques that works only at very high temperature \cite{HLPRA10-2} or only at zero temperature \cite{SadeMeloC11}. Measurements of the compressibility as well as the spin susceptibility in equal-population Fermi gases have been demonstrated in Refs.~\cite{ZwierleinNature11, ZwierleinScience12,ValePRL12} and their generalizations to Fermi superfluids with population imbalance are expected. It is thus of importance to investigate the density and spin response of polarized (population-imbalanced) Fermi superfluids.

Here we present density and spin linear response theories of population-imbalanced Fermi gases in the polarized superfluid regime. It has been emphasized \cite{Mahanbook,OurJLTP13,OurIJMPB} that linear response theories should be fully consistent with several fundamental constraints imposed by the Ward identities, $Q$-limit Ward identities, and sum rules, whose importance has been discussed in Refs.~\cite{OurJLTP13,OurIJMPB}. The Ward identities guarantee gauge invariance, the $Q$-limit Ward identities guarantee the consistency between single-particle thermodynamics and two-particle correlation functions, and sum rules verify conservation laws. We will show that those fundamental constraints can be respected by linear response theories even when the Fermi superfluid is polarized. Here our linear response theories are based on a mean-field description of population-imbalanced Fermi gases in BCS-BEC crossover \cite{OurRoPP,OurAnnPhys}. Going beyond the mean-field descriptions requires the incorporation of pairing fluctuation effects and it has been shown that approximations in those more complicated theories may violate some of the constraints \cite{OurComment,OurC13}. The linear response theories based on the mean-field description presented here serve as a first step in understanding the non-equilibrium behavior of population-imbalanced Fermi superfluids.

The density and spin linear response theories for polarized superfluids presented here are constructed following the consistent-fluctuation-of-the order-parameter (CFOP) approach reviewed in Ref.~\cite{OurJLTP13,OurIJMPB}. This approach has been developed in various forms (but mostly confined to the conventional BCS theory) in previous studies~\cite{Kadanoff61,KulikJLTP81,ZhaPRB95,Arseev} while a generalization to relativistic BCS superfluids is discussed in Ref.~\cite{OurPRD}. As the name implies, the fluctuations of the order parameter must be included to restore gauge invariance of the theory. Those fluctuations, in turn, introduce collective modes and hence the collective mode effects enters into the corresponding response functions. As will be shown shortly, when we follow a similar treatment in the spin channel, the collective mode effects in the spin channel makes the theory qualitatively different from that in the equal-population case.

We caution that the CFOP approach is very different from Nambu's integral-equation approach to a gauge-invariant interacting vertex \cite{Nambu60}, which has been summarized in Schrieffer's book \cite{Schrieffer_book}. As explained in Ref.~\cite{OurJLTP13}, the CFOP approach is computationally manageable and allows for complete expressions for the response functions. The CFOP approach should also be distinguished from the random-phase approximation (RPA) \cite{Mahanbook}, which sums certain bubble diagrams to correct the bare response functions. As pointed out in Refs.~\cite{Mahanbook,OurIJMPB}, RPA or similar approaches have difficulties satisfying the $Q$-limit Ward identity and the associated susceptibility sum rule. This  leads to different expressions from thermodynamics and linear response theories.

Throughout this paper, we follow the convention $c=\hbar=k_B=1$ and use $\sigma$ to denote the spin (or more precisely, the two components) $\uparrow,\downarrow$ with
$\uparrow$ and $\bar{\sigma}$ being the opposite of $\downarrow$ and $\sigma$ respectively. The Hamiltonian of a two-component Fermi gas interacting via contact interactions is
\begin{eqnarray}\label{H00}
H&=&\int d^3\mathbf{x}\sum_{\sigma}\psi^{\dagger}_{\sigma}(\mathbf{x})\Big(\frac{\hat{\mathbf{p}}^2}{2m}-\mu_{\sigma}\Big)\psi_{\sigma}(\mathbf{x})\nonumber\\
&-&g\int
d^3\mathbf{\mathbf{x}}\psi^{\dagger}_{\uparrow}(\mathbf{x})\psi^{\dagger}_{\downarrow}(\mathbf{x})\psi_{\downarrow}(\mathbf{x})\psi_{\uparrow}(\mathbf{x}),
\end{eqnarray}
where $\psi$ and $\psi^{\dagger}$ are the annihilation and creation operators, $\mu_{\sigma}$ is the chemical potential of the component $\sigma$, $g$ is the attractive coupling constant, and $m$ is the fermion mass (assumed equal for the two components). As pointed out by Nambu \cite{Nambu60}, the Hamiltonian has a U(1)$\times$ U(1) symmetry since it is invariant under
\begin{eqnarray}\label{T2}
& &\psi_{\sigma}\rightarrow e^{-i\alpha}\psi_{\sigma},\quad \psi^{\dagger}_{\sigma}\rightarrow e^{i\alpha}\psi^{\dagger}_{\sigma}; \nonumber \\
& &\psi_{\sigma}\rightarrow e^{-iS_{\sigma}\phi}\psi_{\sigma},\quad \psi^{\dagger}_{\sigma}\rightarrow e^{iS_{\sigma}\phi}\psi^{\dagger}_{\sigma}. \label{sz}
\end{eqnarray}
Here $S_{\uparrow,\downarrow}=\pm 1$ and $\alpha$ and $\phi$ are phases for those transformations.
The first transformation is the well-known U(1) symmetry related to electromagnetism (EM) if the particle is charged and perturbed by an EM field. Even for a charge neutral system this symmetry is still present so the system should respect the gauge invariance associated with this symmetry. As a consequence, the mass current is conserved.

The second transformation is the spin rotation around a selected axis (called it the $z-$axis) \cite{Nambu60} and in a two-component atomic Fermi gas this transformation assigns opposite phases to the two components. This symmetry will be associated with the conservation of the spin current. For clarification purposes, we denote the first symmetry by U(1)$_{\textrm{EM}}$ and the second by U(1)$_z$. We also note that the spin susceptibility is
$\partial \delta n/\partial h$, where $\delta n=n_{\uparrow}-n_{\downarrow}$ is the density difference and $h=(\mu_{\uparrow}-\mu_{\downarrow})/2$. This is the analogue of the definition in conventional electron systems, $\partial M/\partial H_{m}$. Here $M$ is the magnetization and $H_{m}$ is the magnetic field.

We present an alternative derivation based on symmetry considerations for linear response. As one will see shortly, this derivation elucidates the importance of Ward identities and other sum rules. This derivation is done by introducing two types of weak external fields such that the two U(1) symmetries becomes ``gauged", i.e. that they become local symmetries. Then by properly including the fluctuations of the order parameter, the gauge invariance can be explicitly maintained. For the density channel, the external field is the weak EM field $A^{\mu}=(\phi,\mathbf{A})$. For the spin channel, the effective external field is $A^{\mu}\equiv(B_z,\mathbf{m})$, where $B_z$ is the $z$ component of the magnetic field and $\mathbf{m}$ is the magnetization.
Ref.~\cite{OurJLTP13} shows how the system, when the external field included, respects the corresponding gauge symmetry by implementing the substitutions $\hat{\mathbf{p}}\rightarrow \hat{\mathbf{p}}-\mathbf{A}$ in the density channel and $\hat{\mathbf{p}}\rightarrow \hat{\mathbf{p}}-S_{\sigma}\mathbf{A}$ in the spin channel.

When the BCS mean-field approximation is considered, the order parameter is given by $\label{Delta}
\Delta(\mathbf{x})=g\langle\psi_{\downarrow}(\mathbf{x})\psi_{\uparrow}(\mathbf{x})\rangle$. The U(1)$_{\textrm{EM}}$ symmetry is spontaneously
broken in the presence of the order parameter below $T_c$. In contrast, the U(1)$_z$ symmetry remains intact. Although these two U(1) symmetries behave differently in the superfluid phase, the frameworks of the two linear response theories are similar. In the equal-population case, the collective mode effects cancel out in the spin channel so the structures of the response functions in the density and spin channels are different. This phenomenon is called a "charge-spin separation" in superfluids \cite{OurJLTP13,OurIJMPB}. Here we will show that, in stark contrast, the collective mode effect plays an important role in the spin channel when the population is imbalanced.

\section{Density Linear Response Theory}\label{Sec.2}
Before deriving the linear response theories, we first explain some terminologies used here.
The Nambu space composes of two-component Nambu spinors \cite{Nambu60}, which corresponds to a two-component representation of fermions.
In the Nambu space one groups $\psi_{{\bf p}\uparrow}$ and $\psi^{\dagger}_{-{\bf p}\downarrow}$ into a two-component column and writes operators as $2\times 2$ matrices, and the propagator and interacting vertices are $2\times 2$ matrices. In the Nambu space, we
will adopt Nambu's convention to name the Ward identity (WI) as the ``generalized Ward identity" (GWI) \cite{Nambu60}. Here we use four-vector notations to compact our expressions for a non-relativistic theory and this has been implemented in many similar circumstances \cite{Schrieffer_book,Mahanbook}.

\subsection{Gauge-invariant vertex and kernel in the density channel}\label{sSec.1}
The density linear response theory is constructed by ``gauging" the U(1)$_{\textrm{EM}}$ symmetry. We define
$\sigma_{\pm}=\frac{1}{2}(\sigma_1\pm i\sigma_2)$ in the Nambu space and introduce
the Nambu-Gorkov spinors
 \begin{equation}\label{Ns}
\Psi_{\mathbf{p}}=\left[\begin{array}{c} \psi_{\mathbf{p}\uparrow} \\
\psi^{\dagger}_{-\mathbf{p}\downarrow}\end{array}\right], \qquad
\Psi^{\dagger}_{\mathbf{p}}=[\psi^{\dagger}_{\mathbf{p}\uparrow},\psi_{-\mathbf{p}\downarrow}].
\end{equation}
The Hamiltonian with the external EM perturbation is given by \begin{eqnarray}\label{HD0}
H&=&\sum_{\mathbf{p}}\Psi^{\dagger}_{\mathbf{p}}\hat{\xi}_{\mathbf{p}}\sigma_3\Psi_{\mathbf{p}}+\sum_{\mathbf{p}\mathbf{q}}
\Psi^{\dagger}_{\mathbf{p}+\mathbf{q}}\big(-\frac{\mathbf{p}+\frac{\mathbf{q}}{2}}{m}\mathbf{A}_{\mathbf{q}}\nonumber\\& &+\Phi_{\mathbf{q}}\sigma_3-\Delta_{\mathbf{q}}\sigma_+-\Delta^*_{-\mathbf{q}}\sigma_-\big)\Psi_{\mathbf{p}},
\end{eqnarray}
where $\hat{\xi}_{\mathbf{p}}=\epsilon_{\mathbf{p}}1_{2\times2}-\hat{\mu}=\textrm{diag}(\xi_{\mathbf{p}\uparrow},\xi_{\mathbf{p}\downarrow})$ with $\epsilon_{\mathbf{p}}=\frac{\mathbf{p}^2}{2m}$ and $\hat{\mu}=\textrm{diag}(\mu_{\uparrow},\mu_{\downarrow})$.
The bare inverse fermion propagator is $\hat{G}^{-1}_0(P)=i\omega_n 1_{2\times 2}-\hat{\xi}_{\mathbf{p}}\sigma_3$ with $P\equiv p^{\mu}=(i\omega_n,\mathbf{p})$ and $\omega_n$ being fermionic Matsubara frequency.

When an external EM field is applied, the order parameter deviates from its equilibrium value $\Delta$, which can be chosen to be real due to the U(1)$_{\textrm{EM}}$ symmetry. The goal of the CFOP approach is to restore the U(1)$_{\textrm{EM}}$ symmetry by finding the corresponding fluctuations of the order parameter.
 We denote the
small perturbation of the order parameter as $\Delta'_{\mathbf{q}}$ so that
$\Delta_{\mathbf{q}}=\Delta+\Delta'_{\mathbf{q}}$. We define
$\Delta_{1\mathbf{q}}=-(\Delta'_{\mathbf{q}}+\Delta^{\prime\ast}_{-\mathbf{q}})/2$
and
$\Delta_{2\mathbf{q}}=-i(\Delta'_{\mathbf{q}}-\Delta^{\prime\ast}_{-\mathbf{q}})/2$ to be the (negative) real and imaginary parts of the perturbation. By introducing the bare EM
interacting vertex
$\hat{\gamma}^{\mu}(P+Q,P)\equiv\hat{\gamma}^{\mu}(\mathbf{p+q},\mathbf{p})=(\sigma_3,\frac{\mathbf{p}+\frac{\mathbf{q}}{2}}{m}1_{2\times 2})$, the Hamiltonian splits into two parts as $H=H_0+H_{\textrm{D}}'$, where
\begin{eqnarray}\label{HD1}
& &H_0=\sum_{\mathbf{p}}\Psi^{\dagger}_{\mathbf{p}}\hat{E}_{\mathbf{p}}\Psi_{\mathbf{p}}, \quad H^{\prime}_{\textrm{D}}=\sum_{\mathbf{p}\mathbf{q}}\Psi^{\dagger}_{\mathbf{p}+\mathbf{q}}\big[\Delta_{1\mathbf{q}}\sigma_1
\nonumber\\
& &\qquad+\Delta_{2\mathbf{q}}\sigma_2+A_{\mu\mathbf{q}}\hat{\gamma}^{\mu}(\mathbf{p}+\mathbf{q},\mathbf{p})\big]\Psi_{\mathbf{p}}.
\end{eqnarray}
Here $\hat{E}_{\mathbf{p}}=\hat{\xi}_{\mathbf{p}}\sigma_3-\Delta\sigma_1$ is the energy operator. Define $\mu=\frac{\mu_{\uparrow}+\mu_{\downarrow}}{2}$,  $h=\frac{\mu_{\uparrow}-\mu_{\downarrow}}{2}$, $\xi_{\mathbf{p}}=\epsilon_{\mathbf{p}}-\mu$ and $E_{\mathbf{p}}=\sqrt{\xi^2_{\mathbf{p}}+\Delta^2}$. It can be verified that $\textrm{det}(\hat{E}_{\mathbf{p}})=-E_{\mathbf{p}\uparrow}E_{\mathbf{p}\downarrow}$, where $E_{\mathbf{p}\uparrow}=E_{\mathbf{p}}-h$ and $E_{\mathbf{p}\downarrow}=E_{\mathbf{p}}+h$ are the quasi-particle energy excitations.

\begin{figure}
 \includegraphics[width=3.4in, clip]{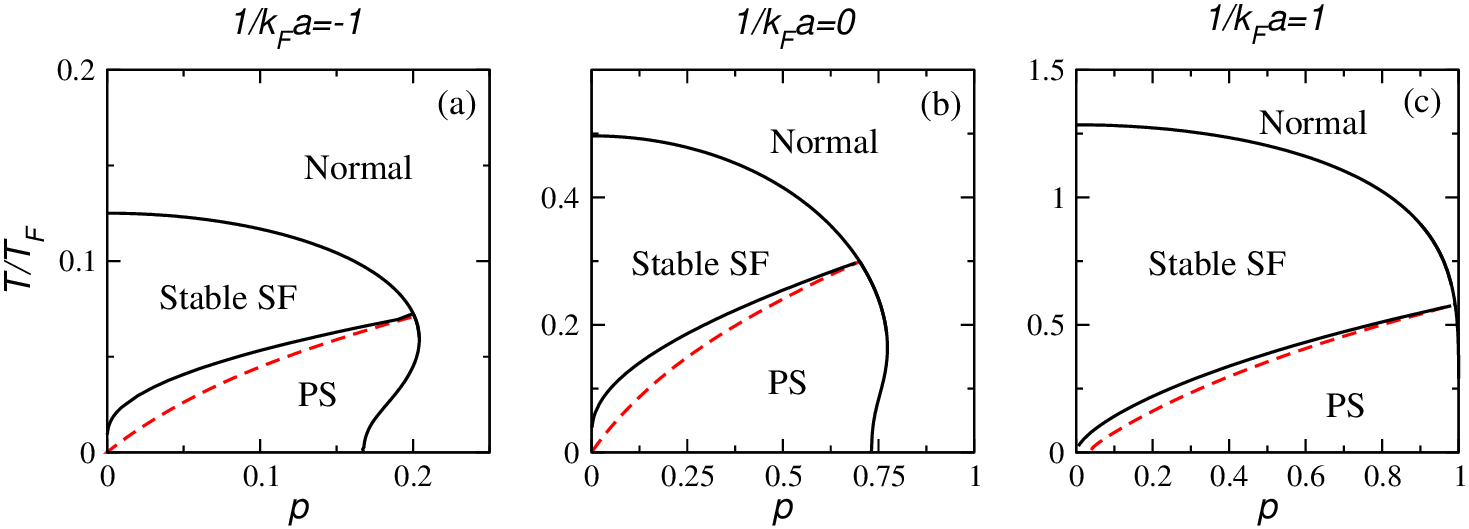}
 \caption{(Color online). Phase diagrams of a homogeneous population-imbalanced Fermi gas in BCS-BEC crossover.  The solid lines separate different phases. The red dashed lines indicate where the polarized superfluid becomes unstable according to $\frac{\partial^2 \Omega}{\partial\Delta^2}$. Here ``SF'' indicates the superfluid phase, ``PS'' means phase separation and ``Normal'' denotes the non-interacting Fermi gas phase. $T_F$ and $k_F$ correspond to the Fermi temperature and Fermi momentum of a noninteracting equal-population Fermi gas with the same total density.}
 \label{fig.0}
\end{figure}

The propagator in the Nambu space is
\begin{eqnarray}\label{G}
\hat{G}(P)=\frac{1}{i\omega_n-\hat{E}_{\mathbf{p}}}=\left(\begin{array}{ll} G_{\uparrow}(P) & F_{\uparrow\downarrow}(P) \\ F_{\downarrow\uparrow}(-P) & -G_{\downarrow}(-P)\end{array}\right),
\end{eqnarray}
which satisfies $\hat{G}^{-1}(P)=\hat{G}^{-1}_0(P)-\hat{\Sigma}(P)$ and $\hat{\Sigma}(P)=-\Delta\sigma_1$ is the self-energy operator in the Nambu space.
The number equation, number difference equation, and gap equation can be extracted from the propagator by $n=\sum_{\sigma}n_{\sigma}=\textrm{Tr}\sum_P\big(\sigma_3\hat{G}(P)\big)$, $\delta n=\textrm{Tr}\sum_P\big(\hat{G}(P)\big)$, and
$\Delta=g\textrm{Tr}\sum_P\big(\sigma_1\hat{G}(P)\big)$. Explicitly,
\begin{eqnarray}\label{NGE}
n&=&\sum_{\mathbf{p}}\Big[1-\frac{\xi_{\mathbf{p}}}{E_{\mathbf{p}}}\big(1-f(E_{\mathbf{p}\uparrow})-f(E_{\mathbf{p}\downarrow})\big)\Big],\nonumber\\
\delta n&=&\sum_{\mathbf{p}}\big(f(E_{\mathbf{p}\uparrow})-f(E_{\mathbf{p}\downarrow})\big),\nonumber\\
\frac{1}{g}&=&-\sum_{\mathbf{p}}\frac{1-f(E_{\mathbf{p}\uparrow})-f(E_{\mathbf{p}\downarrow})}{2E_{\mathbf{p}}}.
\end{eqnarray}
Those equations may be generalized to describe BCS-BEC crossover at the mean-field level.
In BCS-BEC crossover, the coupling constant is related to the dimensionless parameter $1/k_Fa$ via the renormalization $\frac{m}{4\pi a}=\frac{1}{g}+\sum_{\mathbf{p}}\frac{1}{2\epsilon_{\mathbf{p}}}$ \cite{Leggett,largeNcrossover}, where $a$ is the $s$-wave scattering length and $k_F$ is the noninteracting Fermi momentum for the same total number density in the absence
of population imbalance.
The phase diagrams at this mean-field level are shown in Figure~\ref{fig.0} for the BCS ($1/k_F a=-1$), unitary ($1/k_F a=0$), and BEC ($1/k_F a=1$) cases. At low temperature or high polarization, uniform mixed phases are not stable and the system separates into regions of superfluids and normal gases (see Refs.~\cite{ChienPRL,OurRoPP,SheehyRoPP} for more details). Here we focus on the stable polarized superfluid phase and address its response in the density and spin channels.

The Hamiltonian $H'_{\textrm{D}}$ in Eqs.~(\ref{HD1}) including the external EM field can be cast in the form
\begin{eqnarray}\label{HD2}
H_{\textrm{D}}'=\sum_{\mathbf{p}\mathbf{q}}\Psi^{\dagger}_{\mathbf{p}+\mathbf{q}}\hat{\Phi}^T_{\mathbf{q}}\cdot\hat{\Sigma}(\mathbf{p}+\mathbf{q},\mathbf{p})\Psi_{\mathbf{p}},
\end{eqnarray} where $\hat{\mathbf{\Phi}}_{\mathbf{q}}=\big(\Delta_{1\mathbf{q}},\Delta_{2\mathbf{q}},A_{\mu\mathbf{q}}\big)^T$,
$\hat{\mathbf{\Sigma}}(\mathbf{p}+\mathbf{q},\mathbf{p})=\big(\sigma_1,\sigma_2,\hat{\gamma}^{\mu}(\mathbf{p}+\mathbf{q},\mathbf{p})\big)^T$ are defined as the generalized driving potential and generalized interacting vertex, respectively. Hence the contribution of the fluctuations of the order parameter is included in the external driving potential. In the imaginary time formalism, the generalized perturbation of the mass current to the linear order is
\begin{eqnarray}\label{RF0}
\delta \vec{J}_i(\tau,\mathbf{q})&=&\sum_{\mathbf{p}}\langle\Psi^{\dagger}_{\mathbf{p}}(\tau)\hat{\mathbf{\Sigma}}_i(\mathbf{p}+\mathbf{q},\mathbf{p})\Psi_{\mathbf{p}+\mathbf{q}}(\tau)\rangle\nonumber\\
& &+\frac{n}{m}\delta_{i3}h^{\mu\nu}A_{\nu}(\tau,\mathbf{q}),
\end{eqnarray}
where $h^{\mu\nu}=-\eta^{\mu\nu}(1-\eta^{\mu0})$ with $\eta^{\mu\nu}=\textrm{diag}(1,-1,-1,-1)$ being the covariant metric tensor. Here $\delta J_{1,2}$ denotes the perturbation of the real/imaginary part of the order parameter and $\delta J^{\mu}_3$ denotes the perturbed mass current. The density linear response theory can now be cast into a matrix form
\begin{eqnarray}\label{eqn:Q}
& &\delta \vec{J}(Q)=\tensor{Q}(Q)\cdot\hat{\mathbf{\Phi}}(Q)
 \\ &=&\left( \begin{array}{ccc} Q_{11}(Q) &
Q_{12}(Q) & Q^{\nu}_{13}(Q) \\
Q_{21}(Q) & Q_{22}(Q) &
Q^{\nu}_{23}(Q) \\ Q^{\mu}_{31}(Q) &
Q^{\mu}_{32}(Q) &
Q^{\mu\nu}_{33}(Q)+\frac{n}{m}h^{\mu\nu} \end{array}\right)\left(
\begin{array}{ccc} \Delta_{1}(Q) \\ \Delta_{2}(Q) \\A_{\nu}(Q)
\end{array}\right), \nonumber
\end{eqnarray}
where $Q\equiv q^{\mu}=(i\Omega_{l}, \mathbf{q})$ is an external four-momentum and $\Omega_{l}$ is boson Matsubara frequency. After applying Wick's theorem, the density response function is given by
\begin{eqnarray}\label{eqn:RF}
& &Q_{ij}(Q)\nonumber\\
&=&\textrm{Tr}\sum_P
\big[\hat{\Sigma}_{i}(P+Q,P)\hat{G}(P+Q)\hat{\Sigma}_{j}(P,P+Q)\hat{G}(P)\big]\nonumber\\
&=&\textrm{Tr}\sum_{P}
\Big[\hat{\Sigma}_{i}(P+Q,P)\frac{1}{i(\omega_n+\Omega_l)-\hat{E}_{\mathbf{p}+\mathbf{q}}}\nonumber\\
& &\qquad\quad\times\hat{\Sigma}_{j}(P,P+Q)\frac{1}{i\omega_n-\hat{E}_{\mathbf{p}}}\Big],
\end{eqnarray}
where $\sum_P=T\sum_{i\omega_n}\sum_{\mathbf{p}}$. Introducing the $\delta$-function operator $\delta(\epsilon-h-\hat{E}_{\mathbf{p}})$ which can be decomposed as $\delta(\epsilon-h-\hat{E}_{\mathbf{p}})=\hat{u}_{\mathbf{p}}\delta(\epsilon-E_{\mathbf{p}\uparrow})+\hat{v}_{\mathbf{p}}\delta(\epsilon-E_{\mathbf{p}\downarrow})$ with $\hat{u}_{\mathbf{p}}=\frac{1}{2}\big(1+\frac{\hat{E}_{\mathbf{p}}+h}{E_{\mathbf{p}}}\big)$ and $\hat{v}_{\mathbf{p}}=\frac{1}{2}\big(1-\frac{\hat{E}_{\mathbf{p}}+h}{E_{\mathbf{p}}}\big)$, the response function can be further simplified as
%\begin{widetext}
\begin{eqnarray}\label{RFS1}
& &Q_{ij}(Q)=\textrm{Tr}\sum_{\mathbf{p}}\int d\epsilon_1d\epsilon_2\frac{f(\epsilon_2)-f(\epsilon_1)}{i\Omega_l-\epsilon_1+\epsilon_2}\hat{\Sigma}_{i}(\mathbf{p}^+,\mathbf{p}^-)\nonumber\\
&
\times&\delta(\epsilon_1-h-\hat{E}^+_{\mathbf{p}})
\hat{\Sigma}_{j}(\mathbf{p}^-,\mathbf{p}^+)\delta(\epsilon_2-h-\hat{E}^-_{\mathbf{p}}),
\end{eqnarray}
%\end{widetext}
where $\mathbf{p}^{\pm}=\mathbf{p}\pm\frac{\mathbf{q}}{2}$ and $f(\epsilon)=(1+e^{\epsilon/T})^{-1}$ is the Fermi distribution function. From this equality we found the expressions of the response functions summarized in Appendix~\ref{app:b}. We emphasize that knowing the response functions such as $Q^{\mu\nu}$ or $Q_{ij}$ does not lead to the conservation of the mass current straightforwardly. Moreover, proofs of some sum rules (discussed later on) may not be constructed at this level. We will need the gauge-invariant EM response kernel and vertex to complete those tasks.

The gap equation imposes the self-consistent condition $\delta J_{1,2}=-\frac{2}{g}\Delta_{1,2}$.
Applying this relation to Eq.(\ref{eqn:Q}), we can eliminate the dependence of the linear response theory on the external field $\Delta_{1,2}$ and the theory finally reduces to the usual Kubo expression $\delta J^{\mu}_{3}=K^{\mu\nu}A_{\nu}$, where the EM response kernel with the effects of fluctuations of the order parameter included is given by
\begin{widetext}
\begin{eqnarray}\label{dKD} K^{\mu\nu}=\tilde{Q}^{\mu\nu}_{33}+\delta
K^{\mu\nu},\quad \delta
K^{\mu\nu}=-\frac{\tilde{Q}_{11}Q^{\mu}_{32}Q^{\nu}_{23}+\tilde{Q}_{22}Q^{\mu}_{31}Q^{\nu}_{13}-Q_{12}Q^{\mu}_{31}Q^{\nu}_{23}-Q_{21}Q^{\mu}_{32}Q^{\nu}_{13}}{\tilde{Q}_{11}\tilde{Q}_{22}-Q_{12}Q_{21}}.
\end{eqnarray}
\end{widetext}
The poles of $\delta
K^{\mu\nu}$ correspond to collective modes and in the presence of a condensate they include the gapless Nambu-Goldstone mode \cite{KulikJLTP81,Stringari06,OurIJMPB}. Below $T_c$ when the U(1)$_{\textrm{EM}}$ symmetry is broken, it is the contribution associated with the collective modes that restores the gauge invariance. The expression \eqref{dKD} is unlikely to be derived from an RPA-based approach, which usually leads to a kernel of the form $K_{0}/(1+\lambda K_{0})$ \cite{Mahanbook}, where $\lambda$ is a combination of the coupling constant and other factors.

Up to this point we found the expression of the gauge invariant response kernel $K^{\mu\nu}$ without explicitly showing what the gauge invariant EM interacting vertex should be.
From Eq.~(\ref{RFS1}), the non gauge-invariant EM response kernel $\tilde{Q}^{\mu\nu}_{33}$ is expressed as $\tilde{Q}^{\mu\nu}_{33}=\textrm{Tr}\sum_P\big[\hat{\gamma}^{\mu}\hat{G}\hat{\gamma}^{\nu}\hat{G}\big]+\frac{n}{m}h^{\mu\nu}$. The lack of gauge invariance can be remedied by replacing either one of the two bare vertices appearing in $\tilde{Q}^{\mu\nu}_{33}$ by the gauge invariant vertex $\hat{\Gamma}^{\mu}$ \cite{Nambu60}
\begin{eqnarray}\label{Kmn1}
K^{\prime\mu\nu}(Q)&=&\textrm{Tr}\sum_P\big[\hat{\Gamma}^{\mu}(P+Q,P)\hat{G}(P+Q)\nonumber\\
& &\times\hat{\gamma}^{\nu}(P,P+Q)\hat{G}(P)\big]+\frac{n}{m}h^{\mu\nu}.
\end{eqnarray}
The gauge invariant condition $q_{\mu}K^{\prime\mu\nu}=0$ is then equivalent to the GWI for
the vertex
\begin{eqnarray}\label{DGWI}
q_{\mu}\hat{\Gamma}^{\mu}(P+Q,P)=\sigma_3\hat{G}^{-1}(P+Q)-\hat{G}^{-1}(P)\sigma_3.
\end{eqnarray}
The main task of developing a gauge-invariant linear response theory, therefore, is to find a gauge invariant vertex $\hat{\Gamma}^{\mu}$ satisfying
Eq.~\eqref{DGWI}.

Approaches for finding a gauge-invariant vertex are not unique. Nambu \cite{Nambu60} proposed that if the vertex is a solution to an integral equation, it satisfies the GWI \cite{OurJLTP13}. However, it is very hard, if not impossible, to solve this equation exactly. Moreover, it is not known if the $Q$-limit Ward identity is satisfied within this approach. This makes it impractical to verify the compressibility sum rule within this scheme. When considering BCS-BEC crossover, another challenge arises. One has to show how to translate the integral equation from the matrix form in the  Nambu space into a consistent description as dimers form in the BEC regime.

Nevertheless, a fully gauge invariant vertex in the Nambu space can be constructed in the CFOP linear response theory.
Since the interacting vertex should take the form of Eq.~(\ref{Kmn1}) and the CFOP approach leads to Eq.~(\ref{dKD}), the vertex
\begin{eqnarray}\label{FEM}
\hat{\Gamma}^{\mu}(P+Q,P)=\hat{\gamma}^{\mu}(P+Q,P)-\sigma_1\Pi^{\mu}_{1}(Q)-\sigma_2\Pi^{\mu}_{2}(Q), \nonumber\\
\end{eqnarray}
where
\begin{eqnarray}\label{tmp4}
\Pi^{\mu}_{1}=\frac{\left|\begin{array}{cc}Q^{\mu}_{31} & Q_{21}\\ Q^{\mu}_{32} &
\tilde{Q}_{22}\end{array}\right|}{\left|\begin{array}{cc}\tilde{Q}_{11} & Q_{12}\\
Q_{21} & \tilde{Q}_{22}\end{array}\right|}, \mbox{
}\Pi^{\mu}_{2}=\frac{\left|\begin{array}{cc}Q^{\mu}_{32} & Q_{12} \\ Q^{\mu}_{31} &
\tilde{Q}_{11} \end{array}\right|}{\left|\begin{array}{cc}\tilde{Q}_{11} & Q_{12}\\
Q_{21} & \tilde{Q}_{22}\end{array}\right|}
\end{eqnarray}
is a solution.
Since $q_{\mu}\Pi^{\mu}_1(Q)=0$, $q_{\mu}\Pi^{\mu}_2(Q)=-2i\Delta$, it can be verified that
\begin{eqnarray}\label{FGWI}
q_{\mu}\hat{\Gamma}^{\mu}(P+Q,P)&=&q_{\mu}\hat{\gamma}^{\mu}(P+Q,P)+2i\Delta\sigma_2\nonumber\\
&=&\sigma_3\hat{G}^{-1}(P+Q)-\hat{G}^{-1}(P)\sigma_3,
\end{eqnarray}
where Eq.~(\ref{WI0}) has been applied. Therefore the GWI is respected by the interacting vertex from the CFOP scheme.
Using Eqs.~\eqref{dKD}, \eqref{Kmn1}, and \eqref{FGWI}, one can see that the GWI leads to $q_{\mu}K^{\mu\nu}=0$.
Since the conservation of the mass current corresponds to $q_{\mu}\delta J^{\mu}_{3}=0$, which is equivalent to the gauge invariance condition $q_{\mu}K^{\mu\nu}=0$,
this demonstrates the equivalence of gauge invariance, GWI, and conservation of the current mass.

\subsection{Sum rules in the density channel}
The density susceptibility is evaluated from the density response kernel via $\chi_{\rho\rho}=-\frac{1}{\pi}\textrm{Im}K^{00}$, which satisfies the $f$-sum rule
\begin{eqnarray}\label{fsSD}
\int_{-\infty}^{+\infty}d\omega\omega\chi_{\rho\rho}(\omega,\mathbf{q})=n\frac{q^2}{m}.
\end{eqnarray}
Here the analytic continuation $i\Omega_l\rightarrow\omega+i0^+$ has been implemented.
The $f$-sum rule in the density channel is satisfied if the Ward identities of response functions (\ref{GWID}) hold.
The proof is outlined in Appendix~\ref{app:dfsum}.

A nontrivial advantage of working out the interacting vertex, instead of only the response function alone, is that
one can directly verify the compressibility sum rule or density susceptibility sum rule \cite{Kubo_book}
\begin{eqnarray}\label{CSRN0}
\frac{\partial n}{\partial \mu}=-\lim_{\mathbf{q}\rightarrow\mathbf{0}}K^{00}(\omega=0,\mathbf{q}).
\end{eqnarray}
As emphasized in Ref.~\cite{OurJLTP13,OurIJMPB}, this equality has a profound physical implication because it is a connection between the one-particle (thermodynamics) and two-particle (response-function) formalisms. This sum rule is equivalent to the (generalized) $Q$-limit Ward identity \cite{Maebashi09,OurJLTP13,OurIJMPB}, which connects the interacting vertex and the self energy.

Here we show that the gauge-invariance vertex from the CFOP approach satisfies the generalized $Q$-limit Ward identity so
the compressibility sum rule is respected.
In the Nambu space, the generalized $Q$-limit Ward identity is
\begin{equation} \label{DGQWI}
\lim_{\mathbf{q}\rightarrow\mathbf{0}}\hat{\Gamma}^0(P+Q,P)|_{\omega=0}=\frac{\partial \hat{G}^{-1}(P)}{\partial \mu}=\sigma_3-\frac{\partial \hat{\Sigma}(P)}{\partial \mu}.
\end{equation}
Here the analytic continuation $i\Omega_l\rightarrow\omega+i0^+$ has been taken. From Eq.(\ref{FEM}) one can see that \cite{OurJLTP13}
\begin{eqnarray}\label{QWIP1}
& &\lim_{\mathbf{q}\rightarrow\mathbf{0}}\hat{\Gamma}^0(P+Q,P)\Big|_{\omega=0}=\sigma_3-\sigma_1\lim_{\mathbf{q}\rightarrow\mathbf{0}}\Pi_1^0(Q)\Big|_{\omega=0},\nonumber\\
& &\sigma_1\lim_{\mathbf{q}\rightarrow\mathbf{0}}\Pi_1^0(Q)\Big|_{\omega=0}=\sigma_1\frac{\lim_{\mathbf{q}\rightarrow\mathbf{0}}Q^0_{13}(Q)|_{\omega=0}}{\lim_{\mathbf{q}\rightarrow\mathbf{0}}\tilde{Q}_{11}(Q)|_{\omega=0}}\nonumber\\
&=&\frac{\partial \hat{\Sigma}}{\partial \mu},
\end{eqnarray}
where the gap equation has been used to evaluate $\frac{\partial \Delta}{\partial \mu}$.

Thus the CFOP approach in the density channel respects the fundamental constraints given by the Ward identity (and its generalized version in the Nambu space), the $Q$-limit WI (and its generalized version) equivalent the compressibility/density susceptibility sum rule, and the $f$-sum rule.

\section{Spin Linear Response Theory}\label{Sec.3}
The CFOP linear response theory in the spin channel is formulated in a similar way as its density counterpart. The theory is built by gauging the second U(1) symmetry in (\ref{T2}), i.e., the U(1)$_z$ symmetry of the Hamiltonian. The spin linear response theory should respect any gauge transformation of the U(1)$_z$ symmetry and satisfy the analogue constraints discussed in the density channel. The spin $f$-sum rule is a consequence of the conservation of the current associated with the U(1)$_z$ symmetry of the Hamiltonian while the spin compressibility sum rule guarantees the consistency of the expression of the spin susceptibility obtained from thermodynamic and linear response approaches.

A major difference between the two symmetries is that the U(1)$_z$ symmetry of the Hamiltonian is not broken by the BCS order parameter below $T_c$. However, we will see that collective mode effects appear in the spin response functions in the presence of population imbalance. This is very different from the equal-population case \cite{OurJLTP13}, where the collective mode effects from the fluctuations of the order parameter cancel in the spin response functions. As shown in Fig.~\ref{fig.0}, Fermi superfluids with population imbalance may encounter instability towards phase separation and one will see that the collective mode effects in the spin channel contributes significantly as the system approaches the instability.

\subsection{CFOP approach in the spin channel}\label{sSec.3}
In the spin channel, the bare spin interacting vertex is given by $\hat{\gamma}^{\mu}_{\textrm{S}}(P+Q,P)\equiv\hat{\gamma}^{\mu}_{\textrm{S}}(\mathbf{p+q},\mathbf{p})=(1_{2\times 2},\frac{\mathbf{p}+\frac{\mathbf{q}}{2}}{m}\sigma_3)$
in the Nambu space. After separating the order parameter into its equilibrium and perturbed parts, the Hamiltonian is split into $H=H_0+H_{\textrm{S}}'$,  where
\begin{eqnarray}\label{HS1}
& &H^{\prime}_{\textrm{S}}\nonumber\\
&=&\sum_{\mathbf{p}\mathbf{q}}\Psi^{\dagger}_{\mathbf{p}+\mathbf{q}}\big(\Delta_{1\mathbf{q}}\sigma_1+\Delta_{2\mathbf{q}}\sigma_2+A_{\mu\mathbf{q}}\hat{\gamma}_{\textrm{S}}^{\mu}(\mathbf{p}+\mathbf{q},\mathbf{p})\big)\Psi_{\mathbf{p}}\nonumber\\
&=&\sum_{\mathbf{p}\mathbf{q}}\Psi^{\dagger}_{\mathbf{p}+\mathbf{q}}\hat{\Phi}^T_{\mathbf{q}}\cdot\hat{\Sigma}_{\textrm{S}}(\mathbf{p}+\mathbf{q},\mathbf{p})\Psi_{\mathbf{p}}.
\end{eqnarray}
Here $\hat{\mathbf{\Phi}}_{\mathbf{q}}=\big(\Delta_{1\mathbf{q}},\Delta_{2\mathbf{q}},A_{\mu\mathbf{q}}\big)^T$, $\hat{\mathbf{\Sigma}}_{\textrm{S}}(\mathbf{p}+\mathbf{q},\mathbf{p})=\big(\sigma_1,\sigma_2,\hat{\gamma}^{\mu}_{\textrm{S}}(\mathbf{p}+\mathbf{q},\mathbf{p})\big)^T$. Note that the potential $A^{\mu}$ has a different physical meaning from that in the density channel and  we have explained this in Sec.~\ref{Sec.1}. Similar to the density counterpart, the generalized spin response current is
\begin{eqnarray}\label{SRF0}
\delta \vec{J}_{\textrm{S}i}(\tau,\mathbf{q})&=&\sum_{\mathbf{p}}\langle\Psi^{\dagger}_{\mathbf{p}}(\tau)\hat{\mathbf{\Sigma}}_{\textrm{S}i}(\mathbf{p}+\mathbf{q},\mathbf{p})\Psi_{\mathbf{p}+\mathbf{q}}(\tau)\rangle\nonumber\\
& &+\frac{n}{m}\delta_{i3}h^{\mu\nu}A_{\nu}(\tau,\mathbf{q}).
\end{eqnarray}
Here $\delta J^{\mu}_{\textrm{S}3}$ denotes the perturbed spin current. The CFOP linear response is again written in a matrix form
\begin{eqnarray}\label{eqn:SQ}
& &\delta \vec{J}_{\textrm{S}}(Q)=\tensor{Q}_{\textrm{S}}(Q)\cdot\hat{\mathbf{\Phi}}(Q)
 \\ &=&\left( \begin{array}{ccc} Q_{\textrm{S}11}(Q) &
Q_{\textrm{S}12}(Q) & Q^{\nu}_{\textrm{S}13}(Q) \\
Q_{\textrm{S}21}(Q) & Q_{\textrm{S}22}(Q) &
Q^{\nu}_{\textrm{S}23}(Q) \\ Q^{\mu}_{\textrm{S}31}(Q) &
Q^{\mu}_{\textrm{S}32}(Q) &
\tilde{Q}^{\mu\nu}_{\textrm{S}33}(Q) \end{array}\right)\left(
\begin{array}{ccc} \Delta_{1}(Q) \\ \Delta_{2}(Q) \\A_{\nu}(Q)
\end{array}\right), \nonumber
\end{eqnarray}
where $\tilde{Q}^{\mu\nu}_{\textrm{S}33}=Q^{\mu\nu}_{\textrm{S}33}+\frac{n}{m}h^{\mu\nu}$. The spin response functions are given by
\begin{eqnarray}\label{eqn:SRF}
Q_{\textrm{S}ij}(Q)&=&\textrm{Tr}\sum_P
\big[\hat{\Sigma}_{\textrm{S}i}(P+Q,P)\hat{G}(P+Q)\nonumber\\
& &\times\hat{\Sigma}_{\textrm{S}j}(P,P+Q)\hat{G}(P)\big].
\end{eqnarray}

The spin response functions can be evaluated by using similar methods as those used in the density channel and their expressions are listed in Appendix~\ref{app:b}. In the equal-population case with $h=0$, one can verify that $Q^{\mu}_{\textrm{S}13}=Q^{\mu}_{\textrm{S}31}=Q^{\mu}_{\textrm{S}23}=Q^{\mu}_{\textrm{S}32}=0$ \cite{OurJLTP13}. Hence the fluctuations of the order parameter decouple from the spin linear response theory \cite{OurJLTP13}. However, the population-imbalanced case with $h\ne0$ is very different since $h$ may not approach zero even if the number difference between the two species approaches zero, as one can see from the solutions to Eq.~\eqref{NGE}. Therefore the collective mode effects associated with the fluctuations can survive in the presence of population imbalance. Again we caution that it is the spin response kernel and vertex, not those $Q_{\textrm{S}ij}$ listed in Appendix~\ref{app:b}, that allow for straightforward proofs of the conservation of the spin current and sum rules.

Following similar steps as those in the density channel, the spin linear response theory can also be cast in the usual Kubo formalism. Note that $Q_{\textrm{S}11}=Q_{11}$, $Q_{\textrm{S}12}=Q_{12}$, $Q_{\textrm{S}21}=Q_{21}$, and $Q_{\textrm{S}22}=Q_{22}$. Then the self-consistent condition $\delta J_{\textrm{S}1,2}=-\frac{2}{g}\Delta_{1,2}$ in the spin channel is in fact the same as that in the density channel. By applying this and solving Eq.~\eqref{eqn:SQ}, we get $\delta J^{\mu}_{\textrm{S}3}=K^{\mu\nu}_{\textrm{S}33}A_{\nu}$, where
\begin{widetext}
\begin{eqnarray}\label{dKS} K^{\mu\nu}_{\textrm{S}}=\tilde{Q}^{\mu\nu}_{\textrm{S}33}+\delta
K^{\mu\nu}_{\textrm{S}},\quad \delta
K^{\mu\nu}_{\textrm{S}}=-\frac{\tilde{Q}_{\textrm{S}11}Q^{\mu}_{\textrm{S}32}Q^{\nu}_{\textrm{S}23}+\tilde{Q}_{\textrm{S}22}Q^{\mu}_{\textrm{S}31}Q^{\nu}_{\textrm{S}13}-Q_{\textrm{S}12}Q^{\mu}_{\textrm{S}31}Q^{\nu}_{\textrm{S}23}-Q_{\textrm{S}21}Q^{\mu}_{\textrm{S}32}Q^{\nu}_{\textrm{S}13}}{\tilde{Q}_{\textrm{S}11}\tilde{Q}_{\textrm{S}22}-Q_{\textrm{S}12}Q_{\textrm{S}21}}.
\end{eqnarray}
\end{widetext}
Eq.~\eqref{dKS} also shows a different structure from what RPA-based approaches may generate.
The spin response kernel $K^{\mu\nu}_{\textrm{S}}$ may be written formally in a form similar to its counterpart in the density channel:
\begin{eqnarray}\label{KmnS1}
K^{\mu\nu}_{\textrm{S}}(Q)&=&\textrm{Tr}\sum_P\big(\hat{\Gamma}^{\mu}_{\textrm{S}}(P+Q,P)\hat{G}(P+Q)\nonumber\\
& &\times\hat{\gamma}^{\nu}_{\textrm{S}}(P,P+Q)\hat{G}(P)\big)+\frac{n}{m}h^{\mu\nu},
\end{eqnarray}
where $\hat{\Gamma}^{\mu}_{\textrm{S}}(P+Q,P)$ is a U(1)$_z$ gauge invariant interacting vertex satisfying the GWI
\begin{eqnarray}\label{GWI3}
q_{\mu}\hat{\Gamma}^{\mu}_{\textrm{S}}(P+Q,P)=\hat{G}^{-1}(P+Q)-\hat{G}^{-1}(P).
\end{eqnarray}

By comparing the expressions (\ref{KmnS1}) and (\ref{dKS}), the vertex from the CFOP approach is
\begin{equation}\label{FEMS}
\hat{\Gamma}^{\mu}_{\textrm{S}}(P+Q,P)=\hat{\gamma}^{\mu}_{\textrm{S}}(P+Q,P)-\sigma_1\Pi^{\mu}_{\textrm{S}1}(Q)-\sigma_2\Pi^{\mu}_{\textrm{S}2}(Q),
\end{equation}
where
\begin{eqnarray}\label{tmp4S}
\Pi^{\mu}_{\textrm{S}1}=\frac{\left|\begin{array}{cc}Q^{\mu}_{\textrm{S}31} & Q_{\textrm{S}21}\\ Q^{\mu}_{\textrm{S}32} &
\tilde{Q}_{22}\end{array}\right|}{\left|\begin{array}{cc}\tilde{Q}_{\textrm{S}11} & Q_{\textrm{S}12}\\
Q_{\textrm{S}21} & \tilde{Q}_{\textrm{S}22}\end{array}\right|}, \mbox{
}\Pi^{\mu}_{\textrm{S}2}=\frac{\left|\begin{array}{cc}Q^{\mu}_{\textrm{S}32} & Q_{\textrm{S}12} \\ Q^{\mu}_{\textrm{S}31} &
\tilde{Q}_{\textrm{S}11} \end{array}\right|}{\left|\begin{array}{cc}\tilde{Q}_{\textrm{S}11} & Q_{\textrm{S}12}\\
Q_{\textrm{S}21} & \tilde{Q}_{\textrm{S}22}\end{array}\right|}. \end{eqnarray}
A direct calculation shows that the GWI \eqref{GWI3} is indeed satisfied.

\subsection{Constraints in the spin channel}
Here we emphasize the subtlety that the $Q$-limit Ward identity in the spin channel plays a much important role than it does
in the density channel.
One can verify that the bare spin interacting vertex already satisfies the GWI associated with the U(1)$_z$ symmetry in the Nambu space:
\begin{eqnarray}\label{SGWI0}
q_{\mu}\hat{\gamma}^{\mu}_{\textrm{S}}(P+Q,P)&=&\hat{G}^{-1}_0(P+Q)-\hat{G}^{-1}_0(P)\nonumber\\
&=&\hat{G}^{-1}(P+Q)-\hat{G}^{-1}(P).
\end{eqnarray}
This is because the U(1)$_z$ symmetry is not broken by the BCS (pairing) order parameter. Hence the bare spin interacting vertex respects the GWI even without any correction. One must seek further constraints to verify the validity of a spin vertex.

The generalized $Q$-limit Ward identity
\begin{eqnarray}\label{SGQWI}
\lim_{\mathbf{q}\rightarrow\mathbf{0}}\hat{\Gamma}^0_{\textrm{S}}(P+Q,P)|_{\omega=0}=\frac{\partial \hat{G}^{-1}(P)}{\partial h}=1-\frac{\partial \hat{\Sigma}(P)}{\partial h}
\end{eqnarray}
is a sufficient and necessary condition for the spin susceptibility sum rule
\begin{eqnarray}\label{CSRN0S}
\frac{\partial \delta n}{\partial h}=-\lim_{\mathbf{q}\rightarrow\mathbf{0}}K^{00}_{\textrm{S}}(Q)\Big|_{\omega=0}.
\end{eqnarray}
Here the analytic continuation $i\Omega_l\rightarrow\omega+i0^+$ has been applied.
The proof is outlined here.
\begin{eqnarray}\label{CSRNS} \frac{\partial \delta n}{\partial
h}&=&\textrm{Tr}\sum_P\Big(\frac{\partial \hat{G}(P)}{\partial
h}\Big)\nonumber\\
&=&-\textrm{Tr}\sum_P\Big(\hat{G}(P)\big(1-\frac{\partial
\hat{\Sigma}(P)}{\partial h}\big)\hat{G}(P)\Big)\nonumber\\
&=&-\textrm{Tr}\sum_P\Big(\hat{\Gamma}^0_{\textrm{S}}(P,P)\hat{G}(P)\hat{\gamma}^0_{\textrm{S}}(P,P)\hat{G}(P)\Big)\nonumber\\
&=&-\lim_{\mathbf{q}\rightarrow\mathbf{0}}K^{00}_{\textrm{S}}(Q)\Big|_{\omega=0}.
\end{eqnarray}
The importance of this sum rule is that if it is satisfied, one can obtain a consistent expression for the spin
susceptibility from the thermodynamic approach or the linear response theory, whichever is more convenient.

Since $\lim_{\mathbf{q}\rightarrow\mathbf{0}}\hat{\gamma}^0_{\textrm{S}}(P+Q,P)|_{\omega=0}=1$ without the term associated with the self energy, the $Q$-limit WI is not respected by the bare interacting vertex although it satisfies the GWI.
In contrast, the generalized $Q$-limit Ward identity is satisfied by the CFOP vertex:
\begin{eqnarray}\label{QWIP1S}
& &\lim_{\mathbf{q}\rightarrow\mathbf{0}}\hat{\Gamma}^0_{\textrm{S}}(P+Q,P)\Big|_{\omega=0}=1-\sigma_1\lim_{\mathbf{q}\rightarrow\mathbf{0}}\Pi_{\textrm{S}1}^0(Q)\Big|_{\omega=0},\nonumber\\
& &\sigma_1\lim_{\mathbf{q}\rightarrow\mathbf{0}}\Pi_{\textrm{S}1}^0(Q)\Big|_{\omega=0}=\sigma_1\frac{\lim_{\mathbf{q}\rightarrow\mathbf{0}}Q^0_{\textrm{S}13}(Q)|_{\omega=0}}{\lim_{\mathbf{q}\rightarrow\mathbf{0}}\tilde{Q}_{\textrm{S}11}(Q)|_{\omega=0}}\nonumber\\
&=&\frac{\partial \hat{\Sigma}}{\partial h}.
\end{eqnarray}
In the derivation the gap equation has been used to evaluate $\frac{\partial \Delta}{\partial h}$. Furthermore, the GWI \eqref{GWI3} can be verified
by a direct calculation. Thus the interacting vertex from the CFOP theory satisfies both GWI and the $Q$-limit GWI in the spin channel.
We finally come to the important conclusion that the GWI alone can not provide enough constraints of the vertex in the spin channel and the generalized $Q$-limit Ward identity should be used to test the interacting vertex.

We comment briefly on the relation between the bare vertex $\hat{\gamma}^{\mu}_{\textrm{S}}(P+Q,P)$ and the vertex \eqref{FEMS} from the CFOP approach. Although the GWI for the two vertices are slightly different, they both lead to the same Ward identities for spin response functions (Appendix~\ref{app:WIRF}, especially Eq.~\eqref{GWIS}).
This may look counter-intuitive, but one can verify it from direct calculations and there is a physical reason behind it.
From the first two Ward identities in (\ref{GWIS}) we have $q_{\mu}\Pi^{\mu}_{\textrm{S}1,2}(Q)=0$. Hence $\hat{\Gamma}^{\mu}_{\textrm{S}}(P+Q,P)$ indeed differs from the bare vertex $\hat{\gamma}^{\mu}_{\textrm{S}}(P+Q,P)$ only by a U(1)$_z$ gauge transformation.
Since the U(1)$_z$ symmetry is not broken, the GWI cannot discern the difference between the two vertices. One has to resort to the $Q$-limit GWI
to distinguish the two. Moreover, $\delta
K^{\mu\nu}_{\textrm{S}}$  in \eqref{dKS}
has the same denominator as $\delta
K^{\mu\nu}$ does in the density channel. Thus in the presence of population imbalance the collective mode effects enter the spin linear response theory as they did in the density channel.

The generalized Ward identities for spin response functions (\ref{GWIS}) also lead to the $f$-sum rule in the spin channel
\begin{eqnarray}\label{fsS}
\int_{-\infty}^{+\infty}d\omega\omega\chi_{\textrm{SS}}(\omega,\mathbf{q})=n\frac{q^2}{m}.
\end{eqnarray}
Here the spin susceptibility is defined as $\chi_{\textrm{SS}}=-\frac{1}{\pi}\textrm{Im}K^{00}_{\textrm{S}}$, where $K^{00}_{\textrm{S}}$ is the $00$ component of the spin response kernel given by Eq.~(\ref{KmnS1}) with one bare vertex an one corrected vertex $\hat{\Gamma}^0_{\textrm{S}}$. 
The proof of the fact that the CFOP response function satisfies the $f$-sum rule can be constructed by using Eqs.(\ref{GWIS}) and noting that $\mathbf{Q}^0_{\textrm{S}33}=\mathbf{Q}^0_{33}$ and using the identity (\ref{fsl1}) in the Appendix. We caution that if the bare vertex $\hat{\gamma}^0_{\textrm{S}}$ replaces $\hat{\Gamma}^0_{\textrm{S}}$, the $f$-sum rule still holds and a proof is given in Appendix~\ref{app:sfsum}. This implies that the $f$-sum rule can not distinguish the two vertices, either.

In conclusion, we saw that the collective mode effects do play a role in the spin linear response theory for polarized Fermi superfluids. This differs significantly from the unpolarized case, where the collective mode terms cancel \cite{OurIJMPB}. The consistency of the CFOP linear response theory in the spin channel is ensured by the GWI (\ref{GWI3}), the generalized $Q$-limit WI (\ref{SGQWI}) or the spin susceptibility sum rule (\ref{CSRN0S}), and the $f$-sum rule (\ref{fsS}).

We remark that the mean field (BCS) approximation ignores the fluctuations of non-condensed (thermal) pairs so it overestimates the transition temperature in the unitarity and BEC regime \cite{OurRoPP,OurAnnPhys}. When the pairing fluctuations are included, both the self-energy and the interacting vertex must have consistent corrections to ensure the validity of the Ward identities and the $Q$-limit Ward identities. This is beyond the scope of this study and in the following we focus on the linear response theories based on the mean field theory without the non-condensed pairs.

\section{Results and Discussions}
Here we discuss some implications from the linear response theories of polarized Fermi superfluids in BCS-BEC crossover. We are interested in the density and spin susceptibilities, whose expressions are given by
\begin{widetext}
\begin{eqnarray}
\frac{\partial n}{\partial\mu}&=&\sum_{\mathbf{p}}\Big[\frac{\Delta^2}{E^3_{\mathbf{p}}}\big(1-f(E_{\mathbf{p}\uparrow})-f(E_{\mathbf{p}\downarrow})\big)-\frac{\xi^2_{\mathbf{p}}}{E^2_{\mathbf{p}}}\big(\frac{\partial f(E_{\mathbf{p}\uparrow})}{\partial E_{\mathbf{p}\uparrow}}+\frac{\partial f(E_{\mathbf{p}\downarrow})}{\partial E_{\mathbf{p}\downarrow}}\big)\Big]\nonumber\\
&+&\frac{\Big[\sum_{\mathbf{p}}\frac{\xi_{\mathbf{p}}}{E^2_{\mathbf{p}}}\Big(\frac{1-f(E_{\mathbf{p}\uparrow})-f(E_{\mathbf{p}\downarrow})}{E_{\mathbf{p}}}+\frac{\partial f(E_{\mathbf{p}\uparrow})}{\partial E_{\mathbf{p}\uparrow}}+\frac{\partial f(E_{\mathbf{p}\downarrow})}{\partial E_{\mathbf{p}\downarrow}}\Big)\Big]^2}{\sum_{\mathbf{p}}\frac{1}{E^2_{\mathbf{p}}}\Big(\frac{1-f(E_{\mathbf{p}\uparrow})-f(E_{\mathbf{p}\downarrow})}{E_{\mathbf{p}}}+\frac{\partial f(E_{\mathbf{p}\uparrow})}{\partial E_{\mathbf{p}\uparrow}}+\frac{\partial f(E_{\mathbf{p}\downarrow})}{\partial E_{\mathbf{p}\downarrow}}\Big)},\nonumber\\
\frac{\partial \delta n}{\partial h}&=&-\sum_{\mathbf{p}}\Big(\frac{\partial f(E_{\mathbf{p}\uparrow})}{\partial E_{\mathbf{p}\uparrow}}+\frac{\partial f(E_{\mathbf{p}\downarrow})}{\partial E_{\mathbf{p}\downarrow}}\Big)+\frac{\Big[\sum_{\mathbf{p}}\frac{1}{E_{\mathbf{p}}}\Big(\frac{\partial f(E_{\mathbf{p}\uparrow})}{\partial E_{\mathbf{p}\uparrow}}-\frac{\partial f(E_{\mathbf{p}\downarrow})}{\partial E_{\mathbf{p}\downarrow}}\Big)\Big]^2}{\sum_{\mathbf{p}}\frac{1}{E^2_{\mathbf{p}}}\Big(\frac{1-f(E_{\mathbf{p}\uparrow})-f(E_{\mathbf{p}\downarrow})}{E_{\mathbf{p}}}+\frac{\partial f(E_{\mathbf{p}\uparrow})}{\partial E_{\mathbf{p}\uparrow}}+\frac{\partial f(E_{\mathbf{p}\downarrow})}{\partial E_{\mathbf{p}\downarrow}}\Big)}.
\end{eqnarray}
\end{widetext}
Since the CFOP approach respects the $Q$-limit WI in both density and spin channels, one can obtain the same expression either from
the equations of state (Eq.~\eqref{NGE}) or from the response functions. This consistency is not guaranteed in other approaches.
The second term of these two susceptibilities share the same denominator. It is interesting to note that this denominator is proportional to the second partial derivative $\frac{\partial^2 \Omega}{\partial\Delta^2}$, where
\begin{eqnarray}
\Omega=\sum_{\mathbf{p}}(\xi_{\mathbf{p}}-E_{\mathbf{p}})-\sum_{\mathbf{p}\sigma}T\ln(1+e^{-\frac{E_{\mathbf{p}\sigma}}{T}})-\frac{\Delta^2}{g}
\end{eqnarray}
is the thermodynamic potential of a two-component BCS superfluid \cite{OurRoPP}. By a direct calculation,
\begin{eqnarray}
\frac{\partial^2 \Omega}{\partial\Delta^2}&=&\sum_{\mathbf{p}}\frac{\Delta^2}{E^2_{\mathbf{p}}}\Big[\frac{1-f(E_{\mathbf{p}\uparrow})-f(E_{\mathbf{p}\downarrow})}{E_{\mathbf{p}}}  \nonumber \\
& &+\frac{\partial f(E_{\mathbf{p}\uparrow})}{\partial E_{\mathbf{p}\uparrow}}+\frac{\partial f(E_{\mathbf{p}\downarrow})}{\partial E_{\mathbf{p}\downarrow}}\Big].
\end{eqnarray}
$\frac{\partial^2 \Omega}{\partial\Delta^2}>0$ indicates the stability of the homogeneous polarized superfluid phase against phase separation. A generic stability requirement is that the number susceptibility matrix must have positive eigenvalues \cite{Pao}. It can be shown that this requirement is equivalent to the positivity of $\frac{\partial^2 \Omega}{\partial\Delta^2}$ when the gap equation is satisfied \cite{Stability,HePRBS06}. In fact, the two susceptibilities can also be thought of as the two diagonal elements of the full number susceptibility matrix. When the denominator vanishes, the two susceptibilities then diverge.

However, it has been shown that phase separation, which corresponds to an inhomogeneous structure with different phases occupying different regions, already emerges slightly before the denominator changes sign (see Fig.~\ref{fig.0} and Ref.~\cite{Stability}). Thus one may use the vanishing of the denominator as a rough indication that phase separation takes place. Similar to the water-ice transition, the divergence of the spin and density susceptibility as the system approaches the phase separation is physical because the phase separation is associated with a first-order transition where different phases with different densities and polarizations can coexist.
There is another stability condition which is the positivity of the superfluid density. However, at the mean-field level this does not give further unstable boundary \cite{Stability}.

\begin{figure}
\centering
 \includegraphics[width=3.4in, clip]{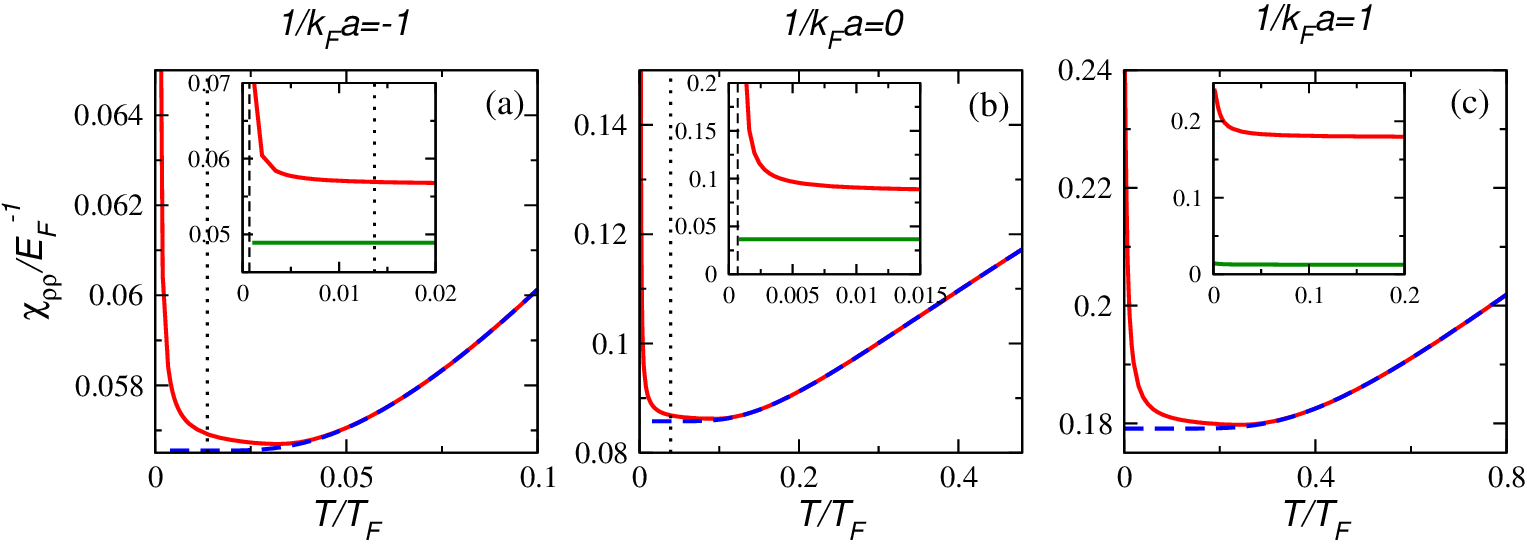}
 \caption{(Color online). Density susceptibility (in units of $E^{-1}_F$) vs. temperature when the polarization is very small ($p=0.001$) from BCS to BEC. The red solid lines indicate the behavior of $dn/d\mu$ and the blue dashed lines denote $dn/d\mu$ for an unpolarized Fermi gas. The insets show the comparison between $dn/d\mu$ and the contribution from its first term (green solid lines) at low temperatures. The black dashed lines indicates where the superfluid becomes unstable according to $\frac{\partial^2 \Omega}{\partial\Delta^2}$ and the black dotted line indicates where phase separation emerges.}
 \label{fig.1}
\end{figure}

\begin{figure}
\centering
 \includegraphics[width=3.4in, clip]{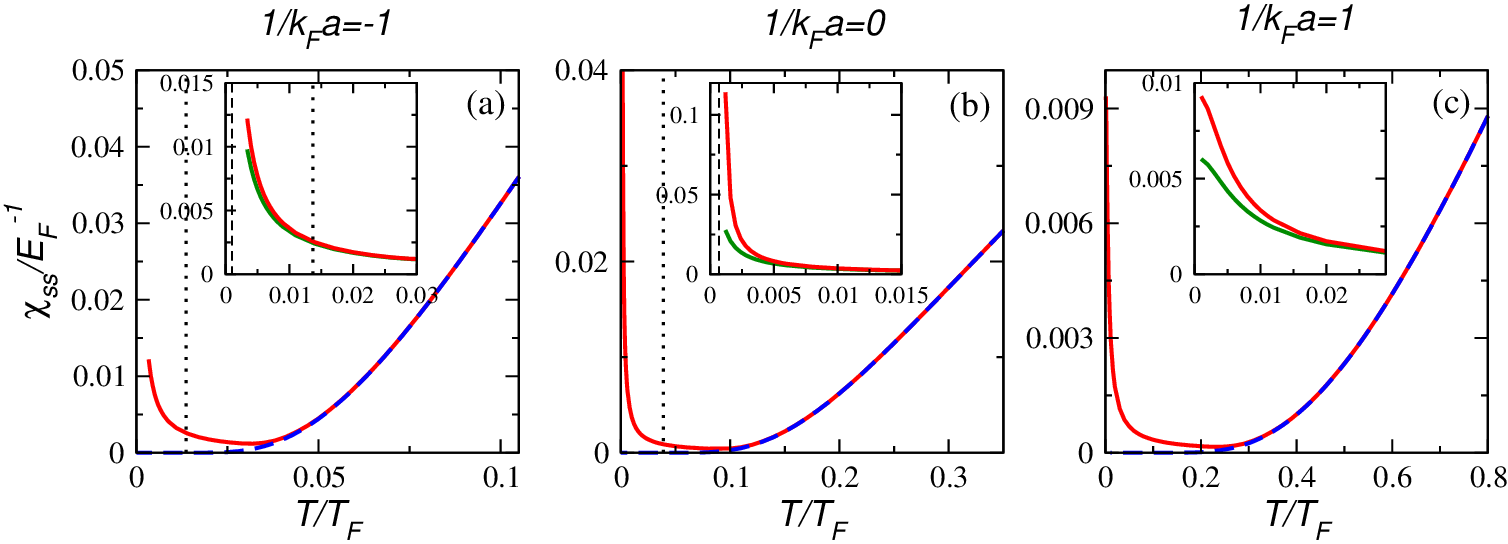}
 \caption{(Color online). Spin susceptibility (in units of $E^{-1}_F$) vs temperature when the polarization is very small ($p=0.001$) from BCS to BEC. The conventions follows those of Fig.\ref{fig.1}.}
 \label{fig.2}
\end{figure}

\begin{figure}
\centering
 \includegraphics[width=3.4in, clip]{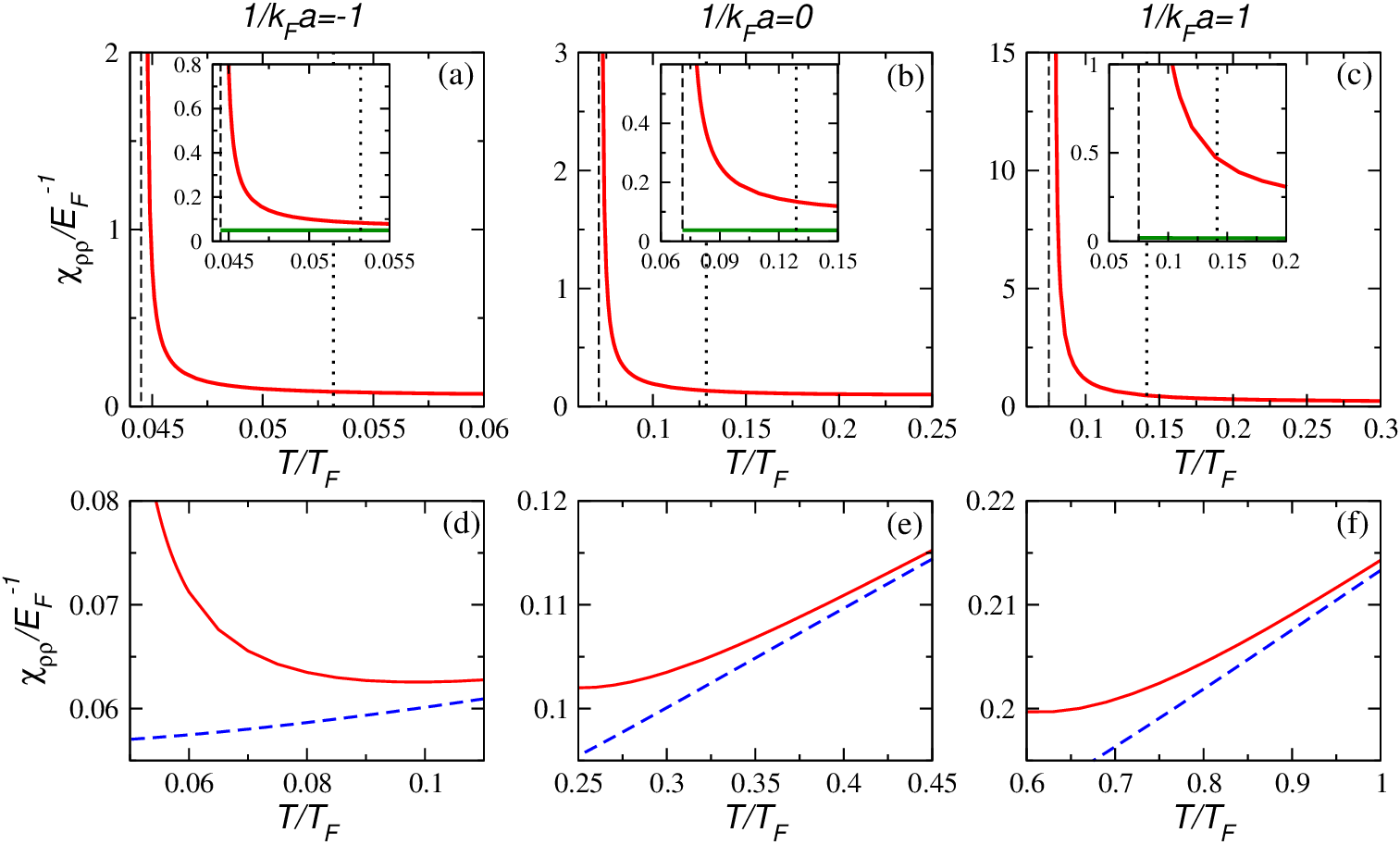}
 \caption{(Color online). Density susceptibility (in units of $E^{-1}_F$) vs. temperature for finite polarization ($p=0.1$) from BCS to BEC. The red solid lines show $dn/d\mu$ and the blue dashed lines denote $dn/d\mu$ for an unpolarized Fermi gas. The green lines denote the contribution from the first term,  the black dashed lines indicate where the superfluid becomes unstable according to $\frac{\partial^2 \Omega}{\partial\Delta^2}$ and the black dotted lines indicate where phase separation emerges. The lower panels ((d), (e), (f)) compare $dn/d\mu$ of polarized and unpolarized Fermi gases at higher  temperatures.}
 \label{fig.3}
\end{figure}

When there is no population imbalance, the density and spin susceptibilities are given by
\begin{eqnarray}\label{eq:eqpop_sus}
\frac{\partial n}{\partial\mu}&=&\sum_{\mathbf{p}}\Big[\frac{\Delta^2}{E^3_{\mathbf{p}}}\big(1-2f(E_{\mathbf{p}})\big)-2\frac{\xi^2_{\mathbf{p}}}{E^2_{\mathbf{p}}}\frac{\partial f(E_{\mathbf{p}})}{\partial E_{\mathbf{p}}}\Big]\nonumber\\
&+&\frac{\Big[\sum_{\mathbf{p}}\frac{\xi_{\mathbf{p}}}{E^2_{\mathbf{p}}}\Big(\frac{1-2f(E_{\mathbf{p}})}{E_{\mathbf{p}}}+2\frac{\partial f(E_{\mathbf{p}})}{\partial E_{\mathbf{p}}}\Big)\Big]^2}{\sum_{\mathbf{p}}\frac{1}{E^2_{\mathbf{p}}}\Big(\frac{1-2f(E_{\mathbf{p}})}{E_{\mathbf{p}}}+2\frac{\partial f(E_{\mathbf{p}})}{\partial E_{\mathbf{p}}}\Big)},\nonumber\\
\frac{\partial \delta n}{\partial h}&=&-2\sum_{\mathbf{p}}\frac{\partial f(E_{\mathbf{p}})}{\partial E_{\mathbf{p}}}.
\end{eqnarray}
Note that $\frac{\partial \delta n}{\partial h}$ does not have the second term shown in the population imbalanced case. This can not be naively proven by setting $h=0$ and hoping that the second term vanishes since $h$ approaches a finite value, not $h\rightarrow 0$, as the number difference approaches zero. After solving Eq.~\eqref{NGE} carefully at $T=0$, one will observe that $h>\Delta$ as long as $p>0$, where $p=\delta n/n$. To find the expressions  \eqref{eq:eqpop_sus}, one has to use the equations of state or linear response theory for the equal population case \cite{OurJLTP13}. This implies that, in the thermodynamic limit, the exactly equal population case is an isolated point and no matter how small the population imbalance is, the spin response  behaves differently when compared to that of the equal population case.
Therefore, the spin linear response theory of unpolarized superfluid should also be viewed as a different theory. The collective mode effects cancel completely in the spin channel in the equal-population case \cite{OurJLTP13}, which has been verified experimentally \cite{ValePRL12}. Moreover, the unpolarized superfluid is always thermodynamically stable in the whole BCS-BEC crossover since there is no gapless quasi-particle excitations due to population imbalance.

\begin{figure}
\centering
 \includegraphics[width=3.5in, clip]{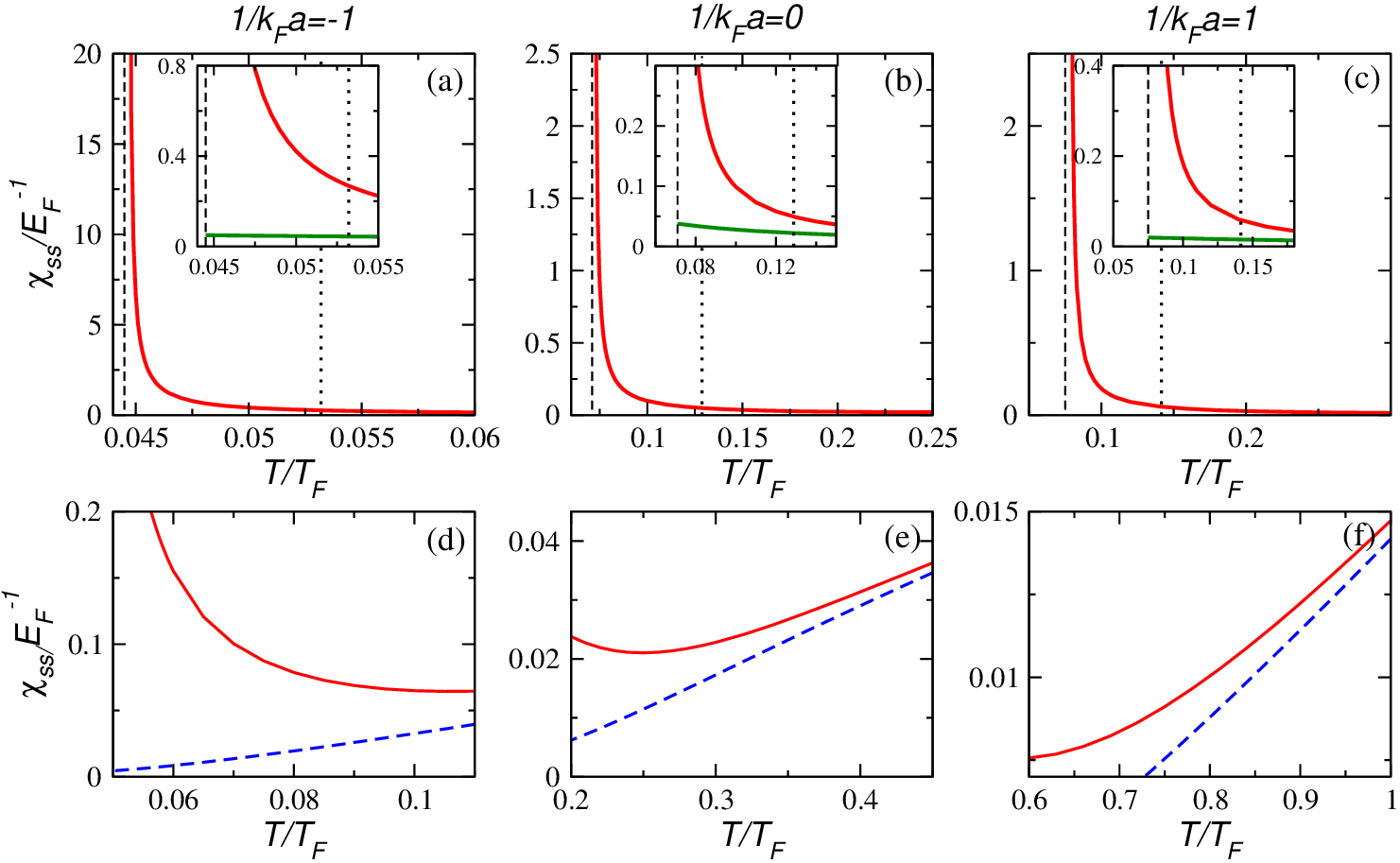}
 \caption{(Color online). Spin susceptibility (in units of $E^{-1}_F$) vs. temperature at relatively high polarization ($p=0.1$) from BCS to BEC. The conventions follow those of Fig.~\ref{fig.3}.}
 \label{fig.4}
\end{figure}
In Figure~\ref{fig.1} we present the density susceptibility (associated with the compressibility) as a function of $T$ for a Fermi gas with small population imbalance in BCS-BEC crossover. The polarization is chosen as $p=0.001$. We compare the behavior of the density susceptibility (red solid lines) to that for an unpolarized Fermi gas (blue dashed lines). One can see that even when the population imbalance is very small, the density susceptibility still shows significantly different behavior at low temperature. At high temperatures, the two curves are indistinguishable. As we explained previously, this difference comes from the second term which is associated with the collective mode effects when $h\ne 0$. Shown in the insets is the comparison between the density susceptibility and the contribution from its first term (shown by the green solid lines) by deliberately dropping the contribution from the collective mode term. When $T$ approaches zero, $\frac{\partial^2 \Omega}{\partial\Delta^2}$ may change sign and if it does (labeled by the black dashed lines), the second term of the density susceptibility dominates the first term and diverges when $\frac{\partial^2 \Omega}{\partial\Delta^2}=0$. Interestingly, in the BEC regime with $1/k_Fa=1$, Fig.~\ref{fig.0} shows that the superfluid phase is stable even at zero temperature for low $p$, but the denominator of the second term or $\frac{\partial^2 \Omega}{\partial\Delta^2}$ is still very small and hence the second term gives a large (but not divergent) contribution.
The black dotted lines indicate that the system enters the phase-separation regime, which are at slightly higher temperatures so the system develops inhomogeneous structures before the susceptibility diverges.

Plotted in Fig.~\ref{fig.2} is the spin susceptibility as a function of $T$ under the same condition of Fig.~\ref{fig.1}. We also show a comparison between the spin susceptibilities of a polarized superfluid (red solid lines) and the unpolarized one (blue dashed lines). At low $T$, the spin susceptibility of polarized Fermi gases shows behavior similar to that of the density susceptibility. It becomes large or even divergent due to the small denominator of the second term. The first term of the spin susceptibility also gives a large contribution at low $T$, as shown in the insets of Fig.~\ref{fig.2}. Again one can see that even when the population imbalance is very small, the polarized Fermi superfluids show qualitatively different behavior when compared to the unpolarized Fermi superfluid.

In Fig.~\ref{fig.3} and Fig.~\ref{fig.4} we increase the population imbalance to $p=0.1$ and show the density and spin susceptibilities as a function of $T$ in BCS-BEC crossover. The black dashed lines indicate the temperature where $\frac{\partial^2 \Omega}{\partial\Delta^2}$ changes sign and the system becomes unstable. Near this temperature, both the density and spin susceptibilities diverge due to the contribution of those terms associated with the collective modes. However, as shown in Fig.~\ref{fig.0}, the phase separation starts to emerge at a slightly higher temperature than that where $\frac{\partial^2 \Omega}{\partial\Delta^2}$ changes sign. Thus in real experiments one should see that the spin and density susceptibilities increase as the system approaches the phase separation regime, but the preemption of the phase separation will cut off the divergent behavior of the susceptibilities. The lower panels of Fig.~\ref{fig.3} and Fig.~\ref{fig.4} ((d), (e), (f)) show the comparisons of the density and spin susceptibilities for polarized (red solid lines) and unpolarized (blue dashed lines) Fermi gases at higher temperatures. One can see that as the polarization increases, the deviation of polarized Fermi gases from the unpolarized one becomes more visible at high temperature.

\section{Conclusion}
We have shown that the density and spin linear response theories of polarized Fermi superfluids based on the CFOP formalism are fully manageable at the mean-field level (with BCS-type approximations) so that we can verify some fundamental constraints explicitly and provide qualitative predictions. The interacting vertices and response functions
satisfy fundamental constraints including the Ward identities, $f$-sum rules, $Q$-limit Ward identities associated with the density and spin susceptibility sum rules. Those constrains guarantee the consistency of the results and gauge invariance associated with the symmetries of the Hamiltonian.

Importantly, we find that collective mode effects from the fluctuations of the order parameter enter both the density and spin channels in the presence of population imbalance. The collective mode effects clearly make the response functions, especially those in the spin channel, different from the  equal-population case. Although the phase separation at low temperature cuts off the divergence in the spin and density response functions, the general trend of increasing susceptibilities as the temperature decreases should be observable.  Future improvements of these linear response theories may include effects from pairing fluctuations or inhomogeneous structures for more detailed descriptions.

We thank Prof. Kathryn Levin for stimulating discussions. Hao Guo thanks the support by National Natural Science Foundation of China (Grants No. 11204032) and Natural Science Foundation of Jiangsu Province, China (SBK201241926). C. C. Chien acknowledges the support of the U.S. Department of Energy through the LANL/LDRD Program.

\appendix
\section{Propagator in the Nambu Space}\label{app:a}
Since
$\hat{\mu}=\mu 1_{2\times 2}+h\sigma_3$, we have $\hat{\xi}_{\mathbf{p}}=\xi_{\mathbf{p}}-h\sigma_3$. The inverse propagator in the mean-field theory is $\hat{G}^{-1}(P)=i\omega_n-\hat{E}_{\mathbf{p}}=i\omega_n-\xi_{\mathbf{p}}\sigma_3+h+\Delta\sigma_1$. Hence
\begin{eqnarray}
\hat{G}(P)&=&\frac{1}{i\omega_n-\xi_{\mathbf{p}}\sigma_3+h+\Delta\sigma_1}\nonumber\\
%&=&\frac{i\omega_n+\xi_{\mathbf{p}}\sigma_3+h-\Delta\sigma_1}{(i\omega_n+h)^2-\xi^2_{\mathbf{p}}-\Delta^2}\nonumber\\
%&=&\frac{\left(\begin{array}{ll} i\omega_n+\xi_{\mathbf{p}\downarrow} & -\Delta \\ -\Delta & i\omega_n-\xi_{\mathbf{p}\uparrow} \end{array}\right)}{(i\omega_n-E_{\mathbf{p}\uparrow})(i\omega_n+E_{\mathbf{p}\downarrow})}\nonumber\\
&=&\left(\begin{array}{ll} G_{\uparrow}(P) & F_{\uparrow\downarrow}(P) \\ F_{\downarrow\uparrow}(-P) & -G_{\downarrow}(-P)\end{array}\right),
\end{eqnarray}
where
\begin{eqnarray}\label{A0}
& &G_{\sigma}(P)=\frac{u^2_{\mathbf{p}}}{i\omega_n-E_{\mathbf{p}\sigma}}+\frac{v^2_{\mathbf{p}}}{i\omega_n+E_{\mathbf{p}\bar{\sigma}}},\nonumber\\
& &F_{\sigma\bar{\sigma}}(P)=-u_{\mathbf{p}}v_{\mathbf{p}}\Big(\frac{1}{i\omega_n-E_{\mathbf{p}\sigma}}-\frac{1}{i\omega_n+E_{\mathbf{p}\bar{\sigma}}}\Big).
\end{eqnarray}
Here $u^2_{\mathbf{p}},v^2_{\mathbf{p}}=\frac{1}{2}(1\pm\frac{\xi_{\mathbf{p}}}{E_{\mathbf{p}}})$.

\section{Ward identities for response functions}\label{app:WIRF}
For the density channel, one can verify that the bare EM interacting vertex and the fermion propagator satisfy
the identity
\begin{equation}\label{WI0}
q_{\mu}\hat{\gamma}^{\mu}(P+Q,P)+2i\Delta\sigma_2 = \sigma_3\hat{G}^{-1}(P+Q)-\hat{G}^{-1}(P)\sigma_3,
\end{equation}
which can verified by using the Green's function in Appendix~\ref{app:a}. By using the response functions in Appendix~\ref{app:b} and defining $\tilde{Q}_{11,22}\equiv \frac{2}{g}+Q_{11,22}$ and $\tilde{Q}^{\mu\nu}_{33}\equiv Q^{\mu\nu}_{33}+\frac{n}{m}h^{\mu\nu}$, Eq.~(\ref{WI0}) is equivalent to
\begin{eqnarray}\label{GWID}
& &q_{\mu}Q_{31}^{\mu}(Q)+2i\Delta
Q_{21}(Q)=0,\nonumber\\
& &q_{\mu}Q_{32}^{\mu}(Q)+2i\Delta
\tilde{Q}_{22}(Q)=0,\nonumber\\
& &q_{\mu}\tilde{Q}_{33}^{\mu\nu}(Q)+2i\Delta
Q^{\nu}_{23}(Q)=0.
\end{eqnarray}
The proof is similar to the equal-population case shown in Ref.~\cite{OurJLTP13}. These are the Ward identities that impose the U(1)$_{\textrm{EM}}$ gauge symmetry on the quantum correlation functions.

For the spin channel, one can verify that the GWI for the bare vertex and the GWI for the CFOP vertex are both equivalent to
\begin{eqnarray}\label{GWIS}
& &q_{\mu}Q_{\textrm{S}31}^{\mu}(Q)=0,\nonumber\\
& &q_{\mu}Q_{\textrm{S}32}^{\mu}(Q)=0,\nonumber\\
& &q_{\mu}\tilde{Q}_{\textrm{S}33}^{\mu\nu}(Q)=0.
\end{eqnarray}
Those are the GWI for spin response functions. This is because the two vertices only differ by an U(1)$_{z}$ gauge transformation, as explained in the main text.
The proofs of Eq.~\eqref{GWIS} are as follows.
By using the GWI (\ref{SGWI0}) and the expression (\ref{eqn:SRF}), we have
\begin{eqnarray}\label{WI1}
& &q_{\mu}Q^{\mu\nu}_{\textrm{S}31}\nonumber\\
&=&\textrm{Tr}\sum_P\Big[q_{\mu}\hat{\gamma}^{\mu}_{\textrm{S}}(P+Q,P)\hat{G}(P+Q)\sigma_1\hat{G}(P)\Big]\nonumber\\
&=&\textrm{Tr}\sum_P\Big[\big(\hat{G}^{-1}(P+Q)-\hat{G}^{-1}(P)\big)\hat{G}(P+Q)\sigma_1\hat{G}(P)\Big]\nonumber\\
&=&\textrm{Tr}\sum_P\Big[\sigma_1\hat{G}(P)\Big]-\textrm{Tr}\sum_P\Big[\hat{G}(P+Q)\sigma_1\Big]\nonumber\\
&=&0,
\end{eqnarray}
where we have changed variables $P\rightarrow P-Q$ for the second term in the third line and used the cyclic property of the trace in the fourth line. If $\sigma_1$ is replaced by $\sigma_2$ in the above derivation, one gets $q_{\mu}Q^{\mu\nu}_{\textrm{S}32}=0$. For the last equality in Eqs.~(\ref{GWIS}), we have
\begin{eqnarray}\label{WI3}
& &q_{\mu}Q^{\mu\nu}_{\textrm{S}33}=\textrm{Tr}\sum_P\Big[\hat{\gamma}^{\nu}_{\textrm{S}}(P+Q,P)\hat{G}(P)\Big]\nonumber\\
&-&\textrm{Tr}\sum_P\Big[\hat{G}(P+Q)\hat{\gamma}^{\nu}_{\textrm{S}}(P,P+Q)\Big]\nonumber\\
&=&\sum_{P}\textrm{Tr}\big([\hat{G}(P)-\hat{G}(P+Q)]\hat{\gamma}^{\nu}_{\textrm{S}}(P,P+Q)\big))\nonumber\\
&=&\sum_{P}\textrm{Tr}\big(\hat{G}(P)[\hat{\gamma}^{\nu}_{\textrm{S}}(P+Q,P)-\hat{\gamma}^{\nu}_{\textrm{S}}(P-Q,P)]\big)\nonumber\\
&=&\frac{q^{\nu}}{m}(1-\eta^{\nu0})\sum_{P}\textrm{Tr}\big(\sigma_3\hat{G}(P)\big)\nonumber\\
&=&-q_{\mu}\frac{n}{m}h^{\mu\nu}.
\end{eqnarray}
Hence we get $q_{\mu}\tilde{Q}^{\mu\nu}_{\textrm{S}33}=0$.

\section{Sketch of the proof of $f$-sum rule in the density channel}\label{app:dfsum}
Here is an outline of the proof of how the CFOP theory satisfies the $f$-sum rule in the density channel.
A simpler version for the unpolarized case can be found in Ref.~\cite{OurJLTP13}.
By using the third equation of the WIs (\ref{GWID}), we have
\begin{widetext}
\begin{eqnarray} \omega K^{00}&=&\mathbf{q}\cdot\mathbf{Q}^0_{33}-2i\Delta
Q^0_{23}-\frac{1}{\tilde{Q}_{11}\tilde{Q}_{22}-Q_{12}Q_{21}}\nonumber\\ &
&\times\big(\tilde{Q}_{11}\mathbf{q}\cdot\mathbf{Q}_{32}Q^0_{23}-2i\Delta\tilde{Q}_{11}\tilde{Q}_{22}Q^0_{23}+\tilde{Q}_{22}\mathbf{q}\cdot\mathbf{Q}_{31}Q^0_{13}-2i\Delta\tilde{Q}_{22}Q_{21}Q^0_{13}\nonumber\\
& &-Q_{12}\mathbf{q}\cdot\mathbf{Q}_{31}Q^0_{23}+2i\Delta Q_{12}Q_{21}Q^0_{23}
-Q_{21}\mathbf{q}\cdot\mathbf{Q}_{32}Q^0_{13}+2i\Delta
Q_{21}\tilde{Q}_{22}Q^0_{13}\big). \end{eqnarray}
\end{widetext}
Note $Q_{12}=-Q_{21}$,
$Q^0_{23}=-Q^0_{32}$, $\mathbf{Q}_{23}=-\mathbf{Q}_{32}$, $Q^0_{13}=Q^0_{31}$ and
$\mathbf{Q}_{13}=\mathbf{Q}_{31}$, we get
\begin{widetext}
 \begin{eqnarray} \omega
K^{00}&=&\mathbf{q}\cdot\mathbf{Q}^0_{33}-\frac{\tilde{Q}_{11}\mathbf{q}\cdot\mathbf{Q}_{32}Q^0_{23}+\tilde{Q}_{22}\mathbf{q}\cdot\mathbf{Q}_{31}Q^0_{13}-Q_{12}\mathbf{q}\cdot\mathbf{Q}_{31}Q^0_{23}-Q_{21}\mathbf{q}\cdot\mathbf{Q}_{32}Q^0_{13}}{\tilde{Q}_{11}\tilde{Q}_{22}-Q_{12}Q_{21}}.
\end{eqnarray}
\end{widetext}
The integral over $\omega$ of the second term is zero since $\tilde{Q}_{11}$, $\tilde{Q}_{22}$, $Q^0_{13}$ and
$\mathbf{Q}_{23}$ are even functions of $\omega$, while $Q_{12}$, $\mathbf{Q}_{13}$
and $Q^0_{23}$ are odd functions of $\omega$. To finish the proof we only need to verify the following identity
\begin{eqnarray}\label{fsl1}
-\int_{-\infty}^{+\infty}d\omega\frac{1}{\pi}\textrm{Im}\big[\mathbf{q}\cdot\mathbf{Q}^0_{33}(\omega,\mathbf{q})\big]=n\frac{q^2}{m}.
\end{eqnarray}

This identity can be verified directly. From the expressions shown in Appendix.\ref{app:b}, we have
\begin{widetext}
\begin{eqnarray}\label{fsE2} &
&-\int_{-\infty}^{+\infty}d\omega\frac{1}{\pi}\textrm{Im}\big[\mathbf{q}\cdot\mathbf{Q}^0_{33}(\omega,\mathbf{q})\big]=\int_{-\infty}^{+\infty}d\omega\sum_{\mathbf{p}}\frac{\mathbf{p}\cdot\mathbf{q}}{2m}\times\nonumber\\
& &\Big\{\big(\frac{\xi^+_{\mathbf{p}}}{E^+_{\mathbf{p}}}-\frac{\xi^-_{\mathbf{p}}}{E^-_{\mathbf{p}}}\big)
\Big(\big[1-f(E^+_{\mathbf{p}\uparrow})-f(E^-_{\mathbf{p}\downarrow})\big]
\delta(\omega-E^+_{\mathbf{p}\uparrow}-E^-_{\mathbf{p}\downarrow})+\big[1-f(E^+_{\mathbf{p}\downarrow})-f(E^-_{\mathbf{p}\uparrow})\big]\delta(\omega+E^+_{\mathbf{p}\downarrow}+E^-_{\mathbf{p}\uparrow})\Big)\nonumber\\
& &-\big(\frac{\xi^+_{\mathbf{p}}}{E^+_{\mathbf{p}}}+\frac{\xi^-_{\mathbf{p}}}{E^-_{\mathbf{p}}}\big)\Big(\big[f(E^+_{\mathbf{p}\uparrow})-f(E^-_{\mathbf{p}\uparrow})\big]
\delta(\omega-E^+_{\mathbf{p}\uparrow}+E^-_{\mathbf{p}\uparrow})+\big[f(E^+_{\mathbf{p}\downarrow})-f(E^-_{\mathbf{p}\downarrow})\big]\delta(\omega+E^+_{\mathbf{p}\downarrow}-E^-_{\mathbf{p}\downarrow})\Big)\Big\}\nonumber\\
&=&\sum_{\mathbf{p}}\frac{\mathbf{p}\cdot\mathbf{q}}{m}\frac{\xi^+_{\mathbf{p}}}{E^+_{\mathbf{p}}}\big(1-f(E^+_{\mathbf{p}\uparrow})-f(E^+_{\mathbf{p}\downarrow})\big)-\sum_{\mathbf{p}}\frac{\mathbf{p}\cdot\mathbf{q}}{m}\frac{\xi^-_{\mathbf{p}}}{E^-_{\mathbf{p}}}\big(1-f(E^-_{\mathbf{p}\uparrow})-f(E^-_{\mathbf{p}\downarrow})\big)\nonumber\\
&=&\sum_{\mathbf{p}}\frac{\mathbf{p}\cdot\mathbf{q}}{m}\big[1-\frac{\xi^-_{\mathbf{p}}}{E^-_{\mathbf{p}}}\big(1-f(E^-_{\mathbf{p}\uparrow})-f(E^-_{\mathbf{p}\downarrow})\big)\big]-\sum_{\mathbf{p}}\frac{\mathbf{p}\cdot\mathbf{q}}{m}\big[1-\frac{\xi^+_{\mathbf{p}}}{E^+_{\mathbf{p}}}\big(1-f(E^+_{\mathbf{p}\uparrow})-f(E^+_{\mathbf{p}\downarrow})\big)\big].
\end{eqnarray}
\end{widetext}
We change variables by
$\mathbf{p}\rightarrow\mathbf{p}+\frac{\mathbf{q}}{2}$ in the first term, and change variables by $\mathbf{p}\rightarrow\mathbf{p}-\frac{\mathbf{q}}{2}$ in the
second term to get
\begin{widetext}
 \begin{eqnarray}\label{fsE3}
-\int_{-\infty}^{+\infty}d\omega\frac{1}{\pi}\textrm{Im}\big[\mathbf{q}\cdot\mathbf{Q}^0_{33}(\omega,\mathbf{q})\big]
=\sum_{\mathbf{p}}\frac{\big[(\mathbf{p}+\frac{\mathbf{q}}{2})-(\mathbf{p}-\frac{\mathbf{q}}{2})\big]\cdot\mathbf{q}}{m}\Big[1-\frac{\xi_{\mathbf{p}}}{E_{\mathbf{p}}}\big(1-f(E_{\mathbf{p}\uparrow})-f(E_{\mathbf{p}\downarrow})\big)\Big]=\frac{q^2}{m}n,
 \end{eqnarray}
 \end{widetext}
where the number equation has been used.

\section{The bare spin vertex in the Nambu space satisfies the $f$-sum rule}\label{app:sfsum}
As pointed out in the main text, the bare spin vertex $\hat{\gamma}^{\mu}_{\textrm{S}}$ already respects the Ward identity (\ref{SGWI0}).
This is different from the density channel, where the bare vertex cannot satisfy the full Ward identity.
This leads to the interesting fact that the $f$-sum rule holds even if we evaluated the spin susceptibility without $\delta K^{00}_{\textrm{S}}$. In other words,  the $f$-sum rule is satisfied even if two bare vertices ($\hat{\gamma}^{\mu}_{\textrm{S}}$) are used in the calculation of the response function. Here is what happens: We define the bare spin susceptibility $\chi^0_{\textrm{S}}=-\frac{1}{\pi}\textrm{Im}\tilde{Q}^{00}_{\textrm{S}33}$, where $\tilde{Q}^{00}_{\textrm{S}33}(Q)=\textrm{Tr}\sum_P\big(\hat{\gamma}^{\mu}_{\textrm{S}}(P+Q,P)\hat{G}(P+Q)\hat{\gamma}^{\nu}_{\textrm{S}}(P,P+Q)\hat{G}(P)\big)$, which includes only the bare vertex. By using the third equality of Eqs.~(\ref{GWIS}) we have
\begin{eqnarray}\label{fsE4}
\int_{-\infty}^{+\infty}d\omega\omega\chi^0_{\textrm{SS}}(\omega,\mathbf{q})&=&\int_{-\infty}^{+\infty}d\omega\omega\tilde{Q}^{00}_{\textrm{S}33}(\omega,\mathbf{q})\nonumber\\
&=&-\int_{-\infty}^{+\infty}d\omega\frac{1}{\pi}\textrm{Im}\big[\mathbf{q}\cdot\mathbf{Q}^0_{\textrm{S}33}(\omega,\mathbf{q})\big]\nonumber\\
&=&\frac{q^2}{m}n.
 \end{eqnarray}

\section{Expressions of Response Functions}\label{app:b}
The response functions in the density channel are given by
\begin{widetext}
\begin{eqnarray}\label{D-Q11}
Q_{11}(\omega,\mathbf{q})&=&\frac{1}{2}\sum_{\mathbf{p}}\Big\{\Big(1+\frac{\xi^+_{\mathbf{p}}\xi^-_{\mathbf{p}}-\Delta^2}{E^+_{\mathbf{p}}E^-_{\mathbf{p}}}\Big)
\Big(\frac{1-f(E^+_{\mathbf{p}\uparrow})-f(E^-_{\mathbf{p}\downarrow})}{\omega-E^+_{\mathbf{p}\uparrow}-E^-_{\mathbf{p}\downarrow}}-\frac{1-f(E^+_{\mathbf{p}\downarrow})-f(E^-_{\mathbf{p}\uparrow})}{\omega+E^+_{\mathbf{p}\downarrow}+E^-_{\mathbf{p}\uparrow}}\Big)
\nonumber\\
& &\qquad-\Big(1-\frac{\xi^+_{\mathbf{p}}\xi^-_{\mathbf{p}}-\Delta^2}{E^+_{\mathbf{p}}E^-_{\mathbf{p}}}\Big)
\Big(\frac{f(E^+_{\mathbf{p}\uparrow})-f(E^-_{\mathbf{p}\uparrow})}{\omega-E^+_{\mathbf{p}\uparrow}+E^-_{\mathbf{p}\uparrow}}-\frac{f(E^+_{\mathbf{p}\downarrow})-f(E^-_{\mathbf{p}\downarrow})}{\omega+E^+_{\mathbf{p}\downarrow}-E^-_{\mathbf{p}\downarrow}}\Big)\Big\},
\end{eqnarray}
\begin{eqnarray}\label{D-Q12}
& &Q_{12}(\omega,\mathbf{q})=Q_{21}(\omega,\mathbf{q})\nonumber\\
&=&-\frac{i}{2}\sum_{\mathbf{p}}\Big\{\Big(\frac{\xi^+_{\mathbf{p}}}{E^+_{\mathbf{p}}}+\frac{\xi^-_{\mathbf{p}}}{E^-_{\mathbf{p}}}\Big)
\Big(\frac{1-f(E^+_{\mathbf{p}\uparrow})-f(E^-_{\mathbf{p}\downarrow})}{\omega-E^+_{\mathbf{p}\uparrow}-E^-_{\mathbf{p}\downarrow}}+\frac{1-f(E^+_{\mathbf{p}\downarrow})-f(E^-_{\mathbf{p}\uparrow})}{\omega+E^+_{\mathbf{p}\downarrow}+E^-_{\mathbf{p}\uparrow}}\Big)
\nonumber\\
& &\qquad\quad-\Big(\frac{\xi^+_{\mathbf{p}}}{E^+_{\mathbf{p}}}-\frac{\xi^-_{\mathbf{p}}}{E^-_{\mathbf{p}}}\Big)
\Big(\frac{f(E^+_{\mathbf{p}\uparrow})-f(E^-_{\mathbf{p}\uparrow})}{\omega-E^+_{\mathbf{p}\uparrow}+E^-_{\mathbf{p}\uparrow}}+\frac{f(E^+_{\mathbf{p}\downarrow})-f(E^-_{\mathbf{p}\downarrow})}{\omega+E^+_{\mathbf{p}\downarrow}-E^-_{\mathbf{p}\downarrow}}\Big)\Big\}, \\
%\end{eqnarray}
%\end{widetext}
%\begin{widetext}
%\begin{eqnarray}\label{D-Q^013}
& &Q^0_{13}(\omega,\mathbf{q})=Q^0_{31}(\omega,\mathbf{q})
=\frac{1}{2}\sum_{\mathbf{p}}\frac{\Delta}{E^+_{\mathbf{p}}E^-_{\mathbf{p}}}(\xi^+_{\mathbf{p}}+\xi^-_{\mathbf{p}})\times\nonumber\\
& &\Big(\frac{1-f(E^+_{\mathbf{p}\uparrow})-f(E^-_{\mathbf{p}\downarrow})}{\omega-E^+_{\mathbf{p}\uparrow}-E^-_{\mathbf{p}\downarrow}}-\frac{1-f(E^+_{\mathbf{p}\downarrow})-f(E^-_{\mathbf{p}\uparrow})}{\omega+E^+_{\mathbf{p}\downarrow}+E^-_{\mathbf{p}\uparrow}}
+
\frac{f(E^+_{\mathbf{p}\uparrow})-f(E^-_{\mathbf{p}\uparrow})}{\omega-E^+_{\mathbf{p}\uparrow}+E^-_{\mathbf{p}\uparrow}}-\frac{f(E^+_{\mathbf{p}\downarrow})-f(E^-_{\mathbf{p}\downarrow})}{\omega+E^+_{\mathbf{p}\downarrow}-E^-_{\mathbf{p}\downarrow}}\Big),
\end{eqnarray}
\begin{eqnarray}\label{D-Q^i13}
& &\mathbf{Q}^i_{13}(\omega,\mathbf{q})=\mathbf{Q}^i_{31}(\omega,\mathbf{q})\nonumber\\
&=&-\frac{\Delta}{2}\sum_{\mathbf{p}}\frac{\mathbf{p}^i}{m}\Big\{\Big(\frac{1}{E^+_{\mathbf{p}}}-\frac{1}{E^-_{\mathbf{p}}}\Big)
\Big(\frac{1-f(E^+_{\mathbf{p}\uparrow})-f(E^-_{\mathbf{p}\downarrow})}{\omega-E^+_{\mathbf{p}\uparrow}-E^-_{\mathbf{p}\downarrow}}+\frac{1-f(E^+_{\mathbf{p}\downarrow})-f(E^-_{\mathbf{p}\uparrow})}{\omega+E^+_{\mathbf{p}\downarrow}+E^-_{\mathbf{p}\uparrow}}\Big)
\nonumber\\
& &\qquad\quad\quad-\Big(\frac{1}{E^+_{\mathbf{p}}}+\frac{1}{E^-_{\mathbf{p}}}\Big)
\Big(\frac{f(E^+_{\mathbf{p}\uparrow})-f(E^-_{\mathbf{p}\uparrow})}{\omega-E^+_{\mathbf{p}\uparrow}+E^-_{\mathbf{p}\uparrow}}+\frac{f(E^+_{\mathbf{p}\downarrow})-f(E^-_{\mathbf{p}\downarrow})}{\omega+E^+_{\mathbf{p}\downarrow}-E^-_{\mathbf{p}\downarrow}}\Big)\Big\}, \\
%\end{eqnarray}
%\end{widetext}
%\begin{widetext}
%\begin{eqnarray}\label{D-Q22}
Q_{22}(\omega,\mathbf{q})&=&\frac{1}{2}\sum_{\mathbf{p}}\Big\{\Big(1+\frac{\xi^+_{\mathbf{p}}\xi^-_{\mathbf{p}}+\Delta^2}{E^+_{\mathbf{p}}E^-_{\mathbf{p}}}\Big)
\Big(\frac{1-f(E^+_{\mathbf{p}\uparrow})-f(E^-_{\mathbf{p}\downarrow})}{\omega-E^+_{\mathbf{p}\uparrow}-E^-_{\mathbf{p}\downarrow}}-\frac{1-f(E^+_{\mathbf{p}\downarrow})-f(E^-_{\mathbf{p}\uparrow})}{\omega+E^+_{\mathbf{p}\downarrow}+E^-_{\mathbf{p}\uparrow}}\Big)
\nonumber\\
& &\qquad-\Big(1-\frac{\xi^+_{\mathbf{p}}\xi^-_{\mathbf{p}}+\Delta^2}{E^+_{\mathbf{p}}E^-_{\mathbf{p}}}\Big)
\Big(\frac{f(E^+_{\mathbf{p}\uparrow})-f(E^-_{\mathbf{p}\uparrow})}{\omega-E^+_{\mathbf{p}\uparrow}+E^-_{\mathbf{p}\uparrow}}-\frac{f(E^+_{\mathbf{p}\downarrow})-f(E^-_{\mathbf{p}\downarrow})}{\omega+E^+_{\mathbf{p}\downarrow}-E^-_{\mathbf{p}\downarrow}}\Big)\Big\},
\end{eqnarray}
\begin{eqnarray}\label{D-Q^023}
& &Q^0_{23}(\omega,\mathbf{q})=-Q^0_{32}(\omega,\mathbf{q})\nonumber\\
&=&\frac{i\Delta}{2}\sum_{\mathbf{p}}\Big\{\Big(\frac{1}{E^+_{\mathbf{p}}}+\frac{1}{E^-_{\mathbf{p}}}\Big)
\Big(\frac{1-f(E^+_{\mathbf{p}\uparrow})-f(E^-_{\mathbf{p}\downarrow})}{\omega-E^+_{\mathbf{p}\uparrow}-E^-_{\mathbf{p}\downarrow}}+\frac{1-f(E^+_{\mathbf{p}\downarrow})-f(E^-_{\mathbf{p}\uparrow})}{\omega+E^+_{\mathbf{p}\downarrow}+E^-_{\mathbf{p}\uparrow}}\Big)
\nonumber\\
& &\qquad\quad-\Big(\frac{1}{E^+_{\mathbf{p}}}-\frac{1}{E^-_{\mathbf{p}}}\Big)
\Big(\frac{f(E^+_{\mathbf{p}\uparrow})-f(E^-_{\mathbf{p}\uparrow})}{\omega-E^+_{\mathbf{p}\uparrow}+E^-_{\mathbf{p}\uparrow}}+\frac{f(E^+_{\mathbf{p}\downarrow})-f(E^-_{\mathbf{p}\downarrow})}{\omega+E^+_{\mathbf{p}\downarrow}-E^-_{\mathbf{p}\downarrow}}\Big)\Big\},
\end{eqnarray}
\begin{eqnarray}\label{D-Q^i23}
& &\mathbf{Q}^i_{23}(\omega,\mathbf{q})=-\mathbf{Q}^i_{32}(\omega,\mathbf{q})
=\frac{i}{2}\sum_{\mathbf{p}}\frac{\mathbf{p}^i}{m}\frac{\Delta}{E^+_{\mathbf{p}}E^-_{\mathbf{p}}}(\xi^+_{\mathbf{p}}-\xi^-_{\mathbf{p}})\times\nonumber\\
& &\Big(\frac{1-f(E^+_{\mathbf{p}\uparrow})-f(E^-_{\mathbf{p}\downarrow})}{\omega-E^+_{\mathbf{p}\uparrow}-E^-_{\mathbf{p}\downarrow}}-\frac{1-f(E^+_{\mathbf{p}\downarrow})-f(E^-_{\mathbf{p}\uparrow})}{\omega+E^+_{\mathbf{p}\downarrow}+E^-_{\mathbf{p}\uparrow}}
+
\frac{f(E^+_{\mathbf{p}\uparrow})-f(E^-_{\mathbf{p}\uparrow})}{\omega-E^+_{\mathbf{p}\uparrow}+E^-_{\mathbf{p}\uparrow}}-\frac{f(E^+_{\mathbf{p}\downarrow})-f(E^-_{\mathbf{p}\downarrow})}{\omega+E^+_{\mathbf{p}\downarrow}-E^-_{\mathbf{p}\downarrow}}\Big),
\end{eqnarray}
\begin{eqnarray}\label{D-Q00}
Q_{33}^{00}(\omega,\mathbf{q})&=&\frac{1}{2}\sum_{\mathbf{p}}\Big\{\Big(1-\frac{\xi^+_{\mathbf{p}}\xi^-_{\mathbf{p}}-\Delta^2}{E^+_{\mathbf{p}}E^-_{\mathbf{p}}}\Big)
\Big(\frac{1-f(E^+_{\mathbf{p}\uparrow})-f(E^-_{\mathbf{p}\downarrow})}{\omega-E^+_{\mathbf{p}\uparrow}-E^-_{\mathbf{p}\downarrow}}-\frac{1-f(E^+_{\mathbf{p}\downarrow})-f(E^-_{\mathbf{p}\uparrow})}{\omega+E^+_{\mathbf{p}\downarrow}+E^-_{\mathbf{p}\uparrow}}\Big)
\nonumber\\
& &\qquad-\Big(1+\frac{\xi^+_{\mathbf{p}}\xi^-_{\mathbf{p}}-\Delta^2}{E^+_{\mathbf{p}}E^-_{\mathbf{p}}}\Big)
\Big(\frac{f(E^+_{\mathbf{p}\uparrow})-f(E^-_{\mathbf{p}\uparrow})}{\omega-E^+_{\mathbf{p}\uparrow}+E^-_{\mathbf{p}\uparrow}}-\frac{f(E^+_{\mathbf{p}\downarrow})-f(E^-_{\mathbf{p}\downarrow})}{\omega+E^+_{\mathbf{p}\downarrow}-E^-_{\mathbf{p}\downarrow}}\Big)\Big\},
\end{eqnarray}
\begin{eqnarray}\label{D-Q0i}
& &\mathbf{Q}_{33}^{0i}(\omega,\mathbf{q})=\mathbf{Q}_{33}^{i0}(\omega,\mathbf{q})\nonumber\\
&=&\frac{1}{2}\sum_{\mathbf{p}}\frac{\mathbf{p}^i}{m}\Big\{\Big(\frac{\xi^+_{\mathbf{p}}}{E^+_{\mathbf{p}}}-\frac{\xi^-_{\mathbf{p}}}{E^-_{\mathbf{p}}}\Big)
\Big(\frac{1-f(E^+_{\mathbf{p}\uparrow})-f(E^-_{\mathbf{p}\downarrow})}{\omega-E^+_{\mathbf{p}\uparrow}-E^-_{\mathbf{p}\downarrow}}+\frac{1-f(E^+_{\mathbf{p}\downarrow})-f(E^-_{\mathbf{p}\uparrow})}{\omega+E^+_{\mathbf{p}\downarrow}+E^-_{\mathbf{p}\uparrow}}\Big)
\nonumber\\
& &\qquad\qquad-\Big(\frac{\xi^+_{\mathbf{p}}}{E^+_{\mathbf{p}}}+\frac{\xi^-_{\mathbf{p}}}{E^-_{\mathbf{p}}}\Big)
\Big(\frac{f(E^+_{\mathbf{p}\uparrow})-f(E^-_{\mathbf{p}\uparrow})}{\omega-E^+_{\mathbf{p}\uparrow}+E^-_{\mathbf{p}\uparrow}}+\frac{f(E^+_{\mathbf{p}\downarrow})-f(E^-_{\mathbf{p}\downarrow})}{\omega+E^+_{\mathbf{p}\downarrow}-E^-_{\mathbf{p}\downarrow}}\Big)\Big\},
\end{eqnarray}
\begin{eqnarray}\label{D-Qij}
\tensor{Q}_{33}^{ij}(\omega,\mathbf{q})&=&\sum_{\mathbf{p}}\frac{\mathbf{p}^i\mathbf{p}^j}{2m^2}\Big\{\Big(1-\frac{\xi^+_{\mathbf{p}}\xi^-_{\mathbf{p}}+\Delta^2}{E^+_{\mathbf{p}}E^-_{\mathbf{p}}}\Big)
\Big(\frac{1-f(E^+_{\mathbf{p}\uparrow})-f(E^-_{\mathbf{p}\downarrow})}{\omega-E^+_{\mathbf{p}\uparrow}-E^-_{\mathbf{p}\downarrow}}-\frac{1-f(E^+_{\mathbf{p}\downarrow})-f(E^-_{\mathbf{p}\uparrow})}{\omega+E^+_{\mathbf{p}\downarrow}+E^-_{\mathbf{p}\uparrow}}\Big)
\nonumber\\
& &\qquad\qquad-\Big(1+\frac{\xi^+_{\mathbf{p}}\xi^-_{\mathbf{p}}+\Delta^2}{E^+_{\mathbf{p}}E^-_{\mathbf{p}}}\Big)
\Big(\frac{f(E^+_{\mathbf{p}\uparrow})-f(E^-_{\mathbf{p}\uparrow})}{\omega-E^+_{\mathbf{p}\uparrow}+E^-_{\mathbf{p}\uparrow}}-\frac{f(E^+_{\mathbf{p}\downarrow})-f(E^-_{\mathbf{p}\downarrow})}{\omega+E^+_{\mathbf{p}\downarrow}-E^-_{\mathbf{p}\downarrow}}\Big)\Big\},
\end{eqnarray}
\end{widetext}

The response functions in the spin channel are given by
\begin{widetext}
\begin{eqnarray}\label{S-Q11}
Q_{\textrm{S}11}(\omega,\mathbf{q})&=&\frac{1}{2}\sum_{\mathbf{p}}\Big\{\Big(1+\frac{\xi^+_{\mathbf{p}}\xi^-_{\mathbf{p}}-\Delta^2}{E^+_{\mathbf{p}}E^-_{\mathbf{p}}}\Big)
\Big(\frac{1-f(E^+_{\mathbf{p}\uparrow})-f(E^-_{\mathbf{p}\downarrow})}{\omega-E^+_{\mathbf{p}\uparrow}-E^-_{\mathbf{p}\downarrow}}-\frac{1-f(E^+_{\mathbf{p}\downarrow})-f(E^-_{\mathbf{p}\uparrow})}{\omega+E^+_{\mathbf{p}\downarrow}+E^-_{\mathbf{p}\uparrow}}\Big)
\nonumber\\
& &\qquad-\Big(1-\frac{\xi^+_{\mathbf{p}}\xi^-_{\mathbf{p}}-\Delta^2}{E^+_{\mathbf{p}}E^-_{\mathbf{p}}}\Big)
\Big(\frac{f(E^+_{\mathbf{p}\uparrow})-f(E^-_{\mathbf{p}\uparrow})}{\omega-E^+_{\mathbf{p}\uparrow}+E^-_{\mathbf{p}\uparrow}}-\frac{f(E^+_{\mathbf{p}\downarrow})-f(E^-_{\mathbf{p}\downarrow})}{\omega+E^+_{\mathbf{p}\downarrow}-E^-_{\mathbf{p}\downarrow}}\Big)\Big\},
\end{eqnarray}
\begin{eqnarray}\label{S-Q12}
& &Q_{\textrm{S}12}(\omega,\mathbf{q})=Q_{\textrm{S}21}(\omega,\mathbf{q})\nonumber\\
&=&-\frac{i}{2}\sum_{\mathbf{p}}\Big\{\Big(\frac{\xi^+_{\mathbf{p}}}{E^+_{\mathbf{p}}}+\frac{\xi^-_{\mathbf{p}}}{E^-_{\mathbf{p}}}\Big)
\Big(\frac{1-f(E^+_{\mathbf{p}\uparrow})-f(E^-_{\mathbf{p}\downarrow})}{\omega-E^+_{\mathbf{p}\uparrow}-E^-_{\mathbf{p}\downarrow}}+\frac{1-f(E^+_{\mathbf{p}\downarrow})-f(E^-_{\mathbf{p}\uparrow})}{\omega+E^+_{\mathbf{p}\downarrow}+E^-_{\mathbf{p}\uparrow}}\Big)
\nonumber\\
& &\qquad\quad-\Big(\frac{\xi^+_{\mathbf{p}}}{E^+_{\mathbf{p}}}-\frac{\xi^-_{\mathbf{p}}}{E^-_{\mathbf{p}}}\Big)
\Big(\frac{f(E^+_{\mathbf{p}\uparrow})-f(E^-_{\mathbf{p}\uparrow})}{\omega-E^+_{\mathbf{p}\uparrow}+E^-_{\mathbf{p}\uparrow}}+\frac{f(E^+_{\mathbf{p}\downarrow})-f(E^-_{\mathbf{p}\downarrow})}{\omega+E^+_{\mathbf{p}\downarrow}-E^-_{\mathbf{p}\downarrow}}\Big)\Big\},
\end{eqnarray}
\begin{eqnarray}\label{S-Q^013}
& &Q^0_{\textrm{S}13}(\omega,\mathbf{q})=Q^0_{\textrm{S}31}(\omega,\mathbf{q})\nonumber\\
&=&-\frac{\Delta}{2}\sum_{\mathbf{p}}\Big\{\Big(\frac{1}{E^+_{\mathbf{p}}}-\frac{1}{E^-_{\mathbf{p}}}\Big)
\Big(\frac{1-f(E^+_{\mathbf{p}\uparrow})-f(E^-_{\mathbf{p}\downarrow})}{\omega-E^+_{\mathbf{p}\uparrow}-E^-_{\mathbf{p}\downarrow}}+\frac{1-f(E^+_{\mathbf{p}\downarrow})-f(E^-_{\mathbf{p}\uparrow})}{\omega+E^+_{\mathbf{p}\downarrow}+E^-_{\mathbf{p}\uparrow}}\Big)
\nonumber\\
& &\qquad\quad-\Big(\frac{1}{E^+_{\mathbf{p}}}+\frac{1}{E^-_{\mathbf{p}}}\Big)
\Big(\frac{f(E^+_{\mathbf{p}\uparrow})-f(E^-_{\mathbf{p}\uparrow})}{\omega-E^+_{\mathbf{p}\uparrow}+E^-_{\mathbf{p}\uparrow}}+\frac{f(E^+_{\mathbf{p}\downarrow})-f(E^-_{\mathbf{p}\downarrow})}{\omega+E^+_{\mathbf{p}\downarrow}-E^-_{\mathbf{p}\downarrow}}\Big)\Big\}, \\
%\end{eqnarray}
%\begin{eqnarray}\label{S-Q^i13}
& &\mathbf{Q}^i_{\textrm{S}13}(\omega,\mathbf{q})=\mathbf{Q}^i_{\textrm{S}31}(\omega,\mathbf{q})
=\frac{1}{2}\sum_{\mathbf{p}}\frac{\mathbf{p}^i}{m}\frac{\Delta}{E^+_{\mathbf{p}}E^-_{\mathbf{p}}}(\xi^+_{\mathbf{p}}+\xi^-_{\mathbf{p}})\times\nonumber\\
& &\Big(\frac{1-f(E^+_{\mathbf{p}\uparrow})-f(E^-_{\mathbf{p}\downarrow})}{\omega-E^+_{\mathbf{p}\uparrow}-E^-_{\mathbf{p}\downarrow}}-\frac{1-f(E^+_{\mathbf{p}\downarrow})-f(E^-_{\mathbf{p}\uparrow})}{\omega+E^+_{\mathbf{p}\downarrow}+E^-_{\mathbf{p}\uparrow}}
+
\frac{f(E^+_{\mathbf{p}\uparrow})-f(E^-_{\mathbf{p}\uparrow})}{\omega-E^+_{\mathbf{p}\uparrow}+E^-_{\mathbf{p}\uparrow}}-\frac{f(E^+_{\mathbf{p}\downarrow})-f(E^-_{\mathbf{p}\downarrow})}{\omega+E^+_{\mathbf{p}\downarrow}-E^-_{\mathbf{p}\downarrow}}\Big), \\
%\end{eqnarray}
%\begin{eqnarray}\label{S-Q22}
Q_{\textrm{S}22}(\omega,\mathbf{q})&=&\frac{1}{2}\sum_{\mathbf{p}}\Big\{\Big(1+\frac{\xi^+_{\mathbf{p}}\xi^-_{\mathbf{p}}+\Delta^2}{E^+_{\mathbf{p}}E^-_{\mathbf{p}}}\Big)
\Big(\frac{1-f(E^+_{\mathbf{p}\uparrow})-f(E^-_{\mathbf{p}\downarrow})}{\omega-E^+_{\mathbf{p}\uparrow}-E^-_{\mathbf{p}\downarrow}}-\frac{1-f(E^+_{\mathbf{p}\downarrow})-f(E^-_{\mathbf{p}\uparrow})}{\omega+E^+_{\mathbf{p}\downarrow}+E^-_{\mathbf{p}\uparrow}}\Big)
\nonumber\\
& &\qquad-\Big(1-\frac{\xi^+_{\mathbf{p}}\xi^-_{\mathbf{p}}+\Delta^2}{E^+_{\mathbf{p}}E^-_{\mathbf{p}}}\Big)
\Big(\frac{f(E^+_{\mathbf{p}\uparrow})-f(E^-_{\mathbf{p}\uparrow})}{\omega-E^+_{\mathbf{p}\uparrow}+E^-_{\mathbf{p}\uparrow}}-\frac{f(E^+_{\mathbf{p}\downarrow})-f(E^-_{\mathbf{p}\downarrow})}{\omega+E^+_{\mathbf{p}\downarrow}-E^-_{\mathbf{p}\downarrow}}\Big)\Big\}, \\
%\end{eqnarray}
%\begin{eqnarray}\label{S-Q^023}
& &Q^0_{\textrm{S}23}(\omega,\mathbf{q})=-Q^0_{\textrm{S}32}(\omega,\mathbf{q})
=\frac{i}{2}\sum_{\mathbf{p}}\frac{\Delta}{E^+_{\mathbf{p}}E^-_{\mathbf{p}}}(\xi^+_{\mathbf{p}}-\xi^-_{\mathbf{p}})\times\nonumber\\
& &\Big(\frac{1-f(E^+_{\mathbf{p}\uparrow})-f(E^-_{\mathbf{p}\downarrow})}{\omega-E^+_{\mathbf{p}\uparrow}-E^-_{\mathbf{p}\downarrow}}-\frac{1-f(E^+_{\mathbf{p}\downarrow})-f(E^-_{\mathbf{p}\uparrow})}{\omega+E^+_{\mathbf{p}\downarrow}+E^-_{\mathbf{p}\uparrow}}
+
\frac{f(E^+_{\mathbf{p}\uparrow})-f(E^-_{\mathbf{p}\uparrow})}{\omega-E^+_{\mathbf{p}\uparrow}+E^-_{\mathbf{p}\uparrow}}-\frac{f(E^+_{\mathbf{p}\downarrow})-f(E^-_{\mathbf{p}\downarrow})}{\omega+E^+_{\mathbf{p}\downarrow}-E^-_{\mathbf{p}\downarrow}}\Big), \\
%\end{eqnarray}
%\begin{eqnarray}\label{S-Q^i23}
& &\mathbf{Q}^i_{\textrm{S}23}(\omega,\mathbf{q})=-\mathbf{Q}^i_{\textrm{S}32}(\omega,\mathbf{q})\nonumber\\
&=&\frac{i\Delta}{2}\sum_{\mathbf{p}}\frac{\mathbf{p}^i}{m}\Big\{\Big(\frac{1}{E^+_{\mathbf{p}}}+\frac{1}{E^-_{\mathbf{p}}}\Big)
\Big(\frac{1-f(E^+_{\mathbf{p}\uparrow})-f(E^-_{\mathbf{p}\downarrow})}{\omega-E^+_{\mathbf{p}\uparrow}-E^-_{\mathbf{p}\downarrow}}+\frac{1-f(E^+_{\mathbf{p}\downarrow})-f(E^-_{\mathbf{p}\uparrow})}{\omega+E^+_{\mathbf{p}\downarrow}+E^-_{\mathbf{p}\uparrow}}\Big)
\nonumber\\
& &\qquad\qquad-\Big(\frac{1}{E^+_{\mathbf{p}}}-\frac{1}{E^-_{\mathbf{p}}}\Big)
\Big(\frac{f(E^+_{\mathbf{p}\uparrow})-f(E^-_{\mathbf{p}\uparrow})}{\omega-E^+_{\mathbf{p}\uparrow}+E^-_{\mathbf{p}\uparrow}}+\frac{f(E^+_{\mathbf{p}\downarrow})-f(E^-_{\mathbf{p}\downarrow})}{\omega+E^+_{\mathbf{p}\downarrow}-E^-_{\mathbf{p}\downarrow}}\Big)\Big\}, \\
%\end{eqnarray}
%\begin{eqnarray}\label{S-Q00}
Q_{\textrm{S}33}^{00}(\omega,\mathbf{q})&=&\frac{1}{2}\sum_{\mathbf{p}}\Big\{\Big(1-\frac{\xi^+_{\mathbf{p}}\xi^-_{\mathbf{p}}+\Delta^2}{E^+_{\mathbf{p}}E^-_{\mathbf{p}}}\Big)
\Big(\frac{1-f(E^+_{\mathbf{p}\uparrow})-f(E^-_{\mathbf{p}\downarrow})}{\omega-E^+_{\mathbf{p}\uparrow}-E^-_{\mathbf{p}\downarrow}}-\frac{1-f(E^+_{\mathbf{p}\downarrow})-f(E^-_{\mathbf{p}\uparrow})}{\omega+E^+_{\mathbf{p}\downarrow}+E^-_{\mathbf{p}\uparrow}}\Big)
\nonumber\\
& &\qquad-\Big(1+\frac{\xi^+_{\mathbf{p}}\xi^-_{\mathbf{p}}+\Delta^2}{E^+_{\mathbf{p}}E^-_{\mathbf{p}}}\Big)
\Big(\frac{f(E^+_{\mathbf{p}\uparrow})-f(E^-_{\mathbf{p}\uparrow})}{\omega-E^+_{\mathbf{p}\uparrow}+E^-_{\mathbf{p}\uparrow}}-\frac{f(E^+_{\mathbf{p}\downarrow})-f(E^-_{\mathbf{p}\downarrow})}{\omega+E^+_{\mathbf{p}\downarrow}-E^-_{\mathbf{p}\downarrow}}\Big)\Big\}, \\
%\end{eqnarray}
%\begin{eqnarray}\label{S-Q0i}
& &\mathbf{Q}_{\textrm{S}33}^{0i}(\omega,\mathbf{q})=\mathbf{Q}_{\textrm{S}33}^{i0}(\omega,\mathbf{q})\nonumber\\
&=&\frac{1}{2}\sum_{\mathbf{p}}\frac{\mathbf{p}^i}{m}\Big\{\Big(\frac{\xi^+_{\mathbf{p}}}{E^+_{\mathbf{p}}}-\frac{\xi^-_{\mathbf{p}}}{E^-_{\mathbf{p}}}\Big)
\Big(\frac{1-f(E^+_{\mathbf{p}\uparrow})-f(E^-_{\mathbf{p}\downarrow})}{\omega-E^+_{\mathbf{p}\uparrow}-E^-_{\mathbf{p}\downarrow}}+\frac{1-f(E^+_{\mathbf{p}\downarrow})-f(E^-_{\mathbf{p}\uparrow})}{\omega+E^+_{\mathbf{p}\downarrow}+E^-_{\mathbf{p}\uparrow}}\Big)
\nonumber\\
& &\qquad\qquad-\Big(\frac{\xi^+_{\mathbf{p}}}{E^+_{\mathbf{p}}}+\frac{\xi^-_{\mathbf{p}}}{E^-_{\mathbf{p}}}\Big)
\Big(\frac{f(E^+_{\mathbf{p}\uparrow})-f(E^-_{\mathbf{p}\uparrow})}{\omega-E^+_{\mathbf{p}\uparrow}+E^-_{\mathbf{p}\uparrow}}+\frac{f(E^+_{\mathbf{p}\downarrow})-f(E^-_{\mathbf{p}\downarrow})}{\omega+E^+_{\mathbf{p}\downarrow}-E^-_{\mathbf{p}\downarrow}}\Big)\Big\},
\end{eqnarray}
\begin{eqnarray}\label{S-Qij}
\tensor{Q}_{\textrm{S}33}^{ij}(\omega,\mathbf{q})&=&\sum_{\mathbf{p}}\frac{\mathbf{p}^i\mathbf{p}^j}{2m^2}\Big\{\Big(1-\frac{\xi^+_{\mathbf{p}}\xi^-_{\mathbf{p}}-\Delta^2}{E^+_{\mathbf{p}}E^-_{\mathbf{p}}}\Big)
\Big(\frac{1-f(E^+_{\mathbf{p}\uparrow})-f(E^-_{\mathbf{p}\downarrow})}{\omega-E^+_{\mathbf{p}\uparrow}-E^-_{\mathbf{p}\downarrow}}-\frac{1-f(E^+_{\mathbf{p}\downarrow})-f(E^-_{\mathbf{p}\uparrow})}{\omega+E^+_{\mathbf{p}\downarrow}+E^-_{\mathbf{p}\uparrow}}\Big)
\nonumber\\
& &\qquad\qquad-\Big(1+\frac{\xi^+_{\mathbf{p}}\xi^-_{\mathbf{p}}-\Delta^2}{E^+_{\mathbf{p}}E^-_{\mathbf{p}}}\Big)
\Big(\frac{f(E^+_{\mathbf{p}\uparrow})-f(E^-_{\mathbf{p}\uparrow})}{\omega-E^+_{\mathbf{p}\uparrow}+E^-_{\mathbf{p}\uparrow}}-\frac{f(E^+_{\mathbf{p}\downarrow})-f(E^-_{\mathbf{p}\downarrow})}{\omega+E^+_{\mathbf{p}\downarrow}-E^-_{\mathbf{p}\downarrow}}\Big)\Big\},
\end{eqnarray}
\end{widetext}

\bibliographystyle{apsrev4-1}
%\bibliography{Review,Review1}

\begin{thebibliography}{42}%
\makeatletter
\providecommand \@ifxundefined [1]{%
 \@ifx{#1\undefined}
}%
\providecommand \@ifnum [1]{%
 \ifnum #1\expandafter \@firstoftwo
 \else \expandafter \@secondoftwo
 \fi
}%
\providecommand \@ifx [1]{%
 \ifx #1\expandafter \@firstoftwo
 \else \expandafter \@secondoftwo
 \fi
}%
\providecommand \natexlab [1]{#1}%
\providecommand \enquote  [1]{``#1''}%
\providecommand \bibnamefont  [1]{#1}%
\providecommand \bibfnamefont [1]{#1}%
\providecommand \citenamefont [1]{#1}%
\providecommand \href@noop [0]{\@secondoftwo}%
\providecommand \href [0]{\begingroup \@sanitize@url \@href}%
\providecommand \@href[1]{\@@startlink{#1}\@@href}%
\providecommand \@@href[1]{\endgroup#1\@@endlink}%
\providecommand \@sanitize@url [0]{\catcode `\\12\catcode `\$12\catcode
  `\&12\catcode `\#12\catcode `\^12\catcode `\_12\catcode `\%12\relax}%
\providecommand \@@startlink[1]{}%
\providecommand \@@endlink[0]{}%
\providecommand \url  [0]{\begingroup\@sanitize@url \@url }%
\providecommand \@url [1]{\endgroup\@href {#1}{\urlprefix }}%
\providecommand \urlprefix  [0]{URL }%
\providecommand \Eprint [0]{\href }%
\@ifxundefined \urlstyle {%
  \providecommand \doi  [0]{\begingroup \@sanitize@url \@doi}%
  \providecommand \@doi [1]{\endgroup \@@startlink {\doibase
  #1}doi:\discretionary {}{}{}#1\@@endlink }%
}{%
  \providecommand \doi  [0]{doi:\discretionary{}{}{}\begingroup
  \urlstyle{rm}\Url }%
}%
\providecommand \doibase [0]{http://dx.doi.org/}%
\providecommand \Doi [0]{\begingroup \@sanitize@url \@Doi }%
\providecommand \@Doi  [1]{\endgroup\@@startlink{\doibase#1}\@@Doi}%
\providecommand \@@Doi [1]{#1\@@endlink}%
\providecommand \selectlanguage [0]{\@gobble}%
\providecommand \bibinfo  [0]{\@secondoftwo}%
\providecommand \bibfield  [0]{\@secondoftwo}%
\providecommand \translation [1]{[#1]}%
\providecommand \BibitemOpen [0]{}%
\providecommand \bibitemStop [0]{}%
\providecommand \bibitemNoStop [0]{.\EOS\space}%
\providecommand \EOS [0]{\spacefactor3000\relax}%
\providecommand \BibitemShut  [1]{\csname bibitem#1\endcsname}%
%</preamble>
\bibitem [{\citenamefont {Giorgini}\ \emph {et~al.}(2008)\citenamefont
  {Giorgini}, \citenamefont {Pitaevskii},\ and\ \citenamefont
  {Stringari}}]{StringariRMP08}%
  \BibitemOpen
  \bibfield  {author} {\bibinfo {author} {\bibfnamefont {S.}~\bibnamefont
  {Giorgini}}, \bibinfo {author} {\bibfnamefont {L.~P.}\ \bibnamefont
  {Pitaevskii}}, \ and\ \bibinfo {author} {\bibfnamefont {S.}~\bibnamefont
  {Stringari}},\ }\href@noop {} {\bibfield  {journal} {\bibinfo  {journal}
  {Rev. Mov. Phys.},\ }\textbf {\bibinfo {volume} {80}},\ \bibinfo {pages}
  {1215} (\bibinfo {year} {2008})}\BibitemShut {NoStop}%
\bibitem [{\citenamefont {Bloch}\ \emph {et~al.}(2008)\citenamefont {Bloch},
  \citenamefont {Dalibard},\ and\ \citenamefont {Zwerger}}]{ZwergerRMP08}%
  \BibitemOpen
  \bibfield  {author} {\bibinfo {author} {\bibfnamefont {I.}~\bibnamefont
  {Bloch}}, \bibinfo {author} {\bibfnamefont {J.}~\bibnamefont {Dalibard}}, \
  and\ \bibinfo {author} {\bibfnamefont {W.}~\bibnamefont {Zwerger}},\
  }\href@noop {} {\bibfield  {journal} {\bibinfo  {journal} {Rev. Mov. Phys.},\
  }\textbf {\bibinfo {volume} {80}},\ \bibinfo {pages} {885} (\bibinfo {year}
  {2008})}\BibitemShut {NoStop}%
\bibitem [{\citenamefont {Pethick}\ and\ \citenamefont
  {Smith}(2008)}]{Pethickbook}%
  \BibitemOpen
  \bibfield  {author} {\bibinfo {author} {\bibfnamefont {C.~J.}\ \bibnamefont
  {Pethick}}\ and\ \bibinfo {author} {\bibfnamefont {H.}~\bibnamefont
  {Smith}},\ }\href@noop {} {\emph {\bibinfo {title} {Bose-Einstein
  Condensation in Dilute Gases}}},\ \bibinfo {edition} {2nd}\ ed.\ (\bibinfo
  {publisher} {Cambridge University Press},\ \bibinfo {address} {Cambridge},\
  \bibinfo {year} {2008})\BibitemShut {NoStop}%
\bibitem [{\citenamefont {Ueda}(2010)}]{Uedabook}%
  \BibitemOpen
  \bibfield  {author} {\bibinfo {author} {\bibfnamefont {M.}~\bibnamefont
  {Ueda}},\ }\href@noop {} {\emph {\bibinfo {title} {Fundamentals and new
  frontiers of Bose-Einstein condensation}}}\ (\bibinfo  {publisher} {World
  scientific},\ \bibinfo {address} {Singapore},\ \bibinfo {year}
  {2010})\BibitemShut {NoStop}%
\bibitem [{\citenamefont {Enss}\ and\ \citenamefont
  {Haussmann}(2012)}]{HaussmannPRL12}%
  \BibitemOpen
  \bibfield  {author} {\bibinfo {author} {\bibfnamefont {T.}~\bibnamefont
  {Enss}}\ and\ \bibinfo {author} {\bibfnamefont {R.}~\bibnamefont
  {Haussmann}},\ }\href@noop {} {\bibfield  {journal} {\bibinfo  {journal}
  {Phys. Rev. Lett.},\ }\textbf {\bibinfo {volume} {109}},\ \bibinfo {pages}
  {195303} (\bibinfo {year} {2012})}\BibitemShut {NoStop}%
\bibitem [{\citenamefont {Palestini}\ \emph {et~al.}(2012)\citenamefont
  {Palestini}, \citenamefont {Pieri},\ and\ \citenamefont
  {Strinati}}]{StrinatiPRL12}%
  \BibitemOpen
  \bibfield  {author} {\bibinfo {author} {\bibfnamefont {F.}~\bibnamefont
  {Palestini}}, \bibinfo {author} {\bibfnamefont {P.}~\bibnamefont {Pieri}}, \
  and\ \bibinfo {author} {\bibfnamefont {G.~C.}\ \bibnamefont {Strinati}},\
  }\href@noop {} {\bibfield  {journal} {\bibinfo  {journal} {Phys. Rev.
  Lett.},\ }\textbf {\bibinfo {volume} {108}},\ \bibinfo {pages} {080401}
  (\bibinfo {year} {2012})}\BibitemShut {NoStop}%
\bibitem [{\citenamefont {Guo}\ \emph {et~al.}(2010)\citenamefont {Guo},
  \citenamefont {Chien},\ and\ \citenamefont {Levin}}]{HaoPRL10}%
  \BibitemOpen
  \bibfield  {author} {\bibinfo {author} {\bibfnamefont {H.}~\bibnamefont
  {Guo}}, \bibinfo {author} {\bibfnamefont {C.~C.}\ \bibnamefont {Chien}}, \
  and\ \bibinfo {author} {\bibfnamefont {K.}~\bibnamefont {Levin}},\
  }\href@noop {} {\bibfield  {journal} {\bibinfo  {journal} {Phys. Rev.
  Lett.},\ }\textbf {\bibinfo {volume} {105}},\ \bibinfo {pages} {120401}
  (\bibinfo {year} {2010})}\BibitemShut {NoStop}%
\bibitem [{\citenamefont {Hoinka}\ \emph {et~al.}(2012)\citenamefont {Hoinka},
  \citenamefont {Lingham}, \citenamefont {Delehaye},\ and\ \citenamefont
  {Vale}}]{ValePRL12}%
  \BibitemOpen
  \bibfield  {author} {\bibinfo {author} {\bibfnamefont {S.}~\bibnamefont
  {Hoinka}}, \bibinfo {author} {\bibfnamefont {M.}~\bibnamefont {Lingham}},
  \bibinfo {author} {\bibfnamefont {M.}~\bibnamefont {Delehaye}}, \ and\
  \bibinfo {author} {\bibfnamefont {C.~J.}\ \bibnamefont {Vale}},\ }\href@noop
  {} {\bibfield  {journal} {\bibinfo  {journal} {Phys. Rev. Lett.},\ }\textbf
  {\bibinfo {volume} {109}},\ \bibinfo {pages} {050403} (\bibinfo {year}
  {2012})}\BibitemShut {NoStop}%
\bibitem [{\citenamefont {Guo}\ \emph {et~al.}(2013){\natexlab{a}}\citenamefont
  {Guo}, \citenamefont {Chien},\ and\ \citenamefont {He}}]{OurJLTP13}%
  \BibitemOpen
  \bibfield  {author} {\bibinfo {author} {\bibfnamefont {H.}~\bibnamefont
  {Guo}}, \bibinfo {author} {\bibfnamefont {C.~C.}\ \bibnamefont {Chien}}, \
  and\ \bibinfo {author} {\bibfnamefont {Y.}~\bibnamefont {He}},\ }\href@noop
  {} {\bibfield  {journal} {\bibinfo  {journal} {J. Low Temp. Phys.},\ }\textbf
  {\bibinfo {volume} {172}},\ \bibinfo {pages} {5} (\bibinfo {year}
  {2013}{\natexlab{a}})}\BibitemShut {NoStop}%
\bibitem [{\citenamefont {Hu}\ \emph {et~al.}(2010){\natexlab{a}}\citenamefont
  {Hu}, \citenamefont {Liu},\ and\ \citenamefont {Drummond}}]{HHEPL10}%
  \BibitemOpen
  \bibfield  {author} {\bibinfo {author} {\bibfnamefont {H.}~\bibnamefont
  {Hu}}, \bibinfo {author} {\bibfnamefont {X.~J.}\ \bibnamefont {Liu}}, \ and\
  \bibinfo {author} {\bibfnamefont {P.~D.}\ \bibnamefont {Drummond}},\
  }\href@noop {} {\bibfield  {journal} {\bibinfo  {journal} {Europhys. Lett.},\
  }\textbf {\bibinfo {volume} {91}},\ \bibinfo {pages} {20005} (\bibinfo {year}
  {2010}{\natexlab{a}})}\BibitemShut {NoStop}%
\bibitem [{\citenamefont {Hu}\ \emph {et~al.}(2010){\natexlab{b}}\citenamefont
  {Hu}, \citenamefont {Liu},\ and\ \citenamefont {Drummond}}]{HLPRA10}%
  \BibitemOpen
  \bibfield  {author} {\bibinfo {author} {\bibfnamefont {H.}~\bibnamefont
  {Hu}}, \bibinfo {author} {\bibfnamefont {X.~J.}\ \bibnamefont {Liu}}, \ and\
  \bibinfo {author} {\bibfnamefont {P.~D.}\ \bibnamefont {Drummond}},\
  }\href@noop {} {\bibfield  {journal} {\bibinfo  {journal} {Phys. Rev. A},\
  }\textbf {\bibinfo {volume} {81}},\ \bibinfo {pages} {033630} (\bibinfo
  {year} {2010}{\natexlab{b}})}\BibitemShut {NoStop}%
\bibitem [{\citenamefont {Hu}\ and\ \citenamefont {Liu}(2012)}]{HLPRA12}%
  \BibitemOpen
  \bibfield  {author} {\bibinfo {author} {\bibfnamefont {H.}~\bibnamefont
  {Hu}}\ and\ \bibinfo {author} {\bibfnamefont {X.~J.}\ \bibnamefont {Liu}},\
  }\href@noop {} {\bibfield  {journal} {\bibinfo  {journal} {Phys. Rev. A},\
  }\textbf {\bibinfo {volume} {85}},\ \bibinfo {pages} {023612} (\bibinfo
  {year} {2012})}\BibitemShut {NoStop}%
\bibitem [{spi()}]{spin_note}%
  \BibitemOpen
  \href@noop {} {}\bibinfo {note} {In ultracold atomic Fermi gases, the two
  components are usually from two hyperfine states of the same species, but
  following the convention of electrons we call the two components spin up and
  spin down.}\BibitemShut {Stop}%
\bibitem [{\citenamefont {Zwierlein}\ \emph {et~al.}(2006)\citenamefont
  {Zwierlein}, \citenamefont {Schirotzek}, \citenamefont {Schunck},\ and\
  \citenamefont {Ketterle}}]{ZSSK06}%
  \BibitemOpen
  \bibfield  {author} {\bibinfo {author} {\bibfnamefont {M.~W.}\ \bibnamefont
  {Zwierlein}}, \bibinfo {author} {\bibfnamefont {A.}~\bibnamefont
  {Schirotzek}}, \bibinfo {author} {\bibfnamefont {C.~H.}\ \bibnamefont
  {Schunck}}, \ and\ \bibinfo {author} {\bibfnamefont {W.}~\bibnamefont
  {Ketterle}},\ }\href@noop {} {\bibfield  {journal} {\bibinfo  {journal}
  {Science},\ }\textbf {\bibinfo {volume} {311}},\ \bibinfo {pages} {492}
  (\bibinfo {year} {2006})}\BibitemShut {NoStop}%
\bibitem [{\citenamefont {Partridge}\ \emph {et~al.}(2006)\citenamefont
  {Partridge}, \citenamefont {Li}, \citenamefont {Kamar}, \citenamefont
  {Liao},\ and\ \citenamefont {Hulet}}]{Rice1}%
  \BibitemOpen
  \bibfield  {author} {\bibinfo {author} {\bibfnamefont {G.~B.}\ \bibnamefont
  {Partridge}}, \bibinfo {author} {\bibfnamefont {W.}~\bibnamefont {Li}},
  \bibinfo {author} {\bibfnamefont {R.~I.}\ \bibnamefont {Kamar}}, \bibinfo
  {author} {\bibfnamefont {Y.~A.}\ \bibnamefont {Liao}}, \ and\ \bibinfo
  {author} {\bibfnamefont {R.~G.}\ \bibnamefont {Hulet}},\ }\href@noop {}
  {\bibfield  {journal} {\bibinfo  {journal} {Science},\ }\textbf {\bibinfo
  {volume} {311}},\ \bibinfo {pages} {503} (\bibinfo {year}
  {2006})}\BibitemShut {NoStop}%
\bibitem [{\citenamefont {Chien}\ \emph {et~al.}(2007)\citenamefont {Chien},
  \citenamefont {Chen}, \citenamefont {He},\ and\ \citenamefont
  {Levin}}]{ChienPRL}%
  \BibitemOpen
  \bibfield  {author} {\bibinfo {author} {\bibfnamefont {C.-C.}\ \bibnamefont
  {Chien}}, \bibinfo {author} {\bibfnamefont {Q.~J.}\ \bibnamefont {Chen}},
  \bibinfo {author} {\bibfnamefont {Y.}~\bibnamefont {He}}, \ and\ \bibinfo
  {author} {\bibfnamefont {K.}~\bibnamefont {Levin}},\ }\href@noop {}
  {\bibfield  {journal} {\bibinfo  {journal} {Phys. Rev. Lett.},\ }\textbf
  {\bibinfo {volume} {98}},\ \bibinfo {pages} {110404} (\bibinfo {year}
  {2007})}\BibitemShut {NoStop}%
\bibitem [{\citenamefont {Radzihovsky}\ and\ \citenamefont
  {Sheehy}(2010)}]{SheehyRoPP}%
  \BibitemOpen
  \bibfield  {author} {\bibinfo {author} {\bibfnamefont {L.}~\bibnamefont
  {Radzihovsky}}\ and\ \bibinfo {author} {\bibfnamefont {D.~E.}\ \bibnamefont
  {Sheehy}},\ }\href@noop {} {\bibfield  {journal} {\bibinfo  {journal} {Rep.
  Prog. Phys.},\ }\textbf {\bibinfo {volume} {73}},\ \bibinfo {pages} {076501}
  (\bibinfo {year} {2010})}\BibitemShut {NoStop}%
\bibitem [{\citenamefont {Guo}\ \emph {et~al.}(2013){\natexlab{b}}\citenamefont
  {Guo}, \citenamefont {Chien}, \citenamefont {He},\ and\ \citenamefont
  {Levin}}]{OurIJMPB}%
  \BibitemOpen
  \bibfield  {author} {\bibinfo {author} {\bibfnamefont {H.}~\bibnamefont
  {Guo}}, \bibinfo {author} {\bibfnamefont {C.~C.}\ \bibnamefont {Chien}},
  \bibinfo {author} {\bibfnamefont {Y.}~\bibnamefont {He}}, \ and\ \bibinfo
  {author} {\bibfnamefont {K.}~\bibnamefont {Levin}},\ }\href@noop {}
  {\bibfield  {journal} {\bibinfo  {journal} {Int. J. Mod. Phys. B},\ }\textbf
  {\bibinfo {volume} {27}},\ \bibinfo {pages} {1330010} (\bibinfo {year}
  {2013}{\natexlab{b}})}\BibitemShut {NoStop}%
\bibitem [{\citenamefont {Liu}\ and\ \citenamefont {Hu}(2010)}]{HLPRA10-2}%
  \BibitemOpen
  \bibfield  {author} {\bibinfo {author} {\bibfnamefont {X.~J.}\ \bibnamefont
  {Liu}}\ and\ \bibinfo {author} {\bibfnamefont {H.}~\bibnamefont {Hu}},\
  }\href@noop {} {\bibfield  {journal} {\bibinfo  {journal} {Phys. Rev. A},\
  }\textbf {\bibinfo {volume} {82}},\ \bibinfo {pages} {043626} (\bibinfo
  {year} {2010})}\BibitemShut {NoStop}%
\bibitem [{\citenamefont {Seo}\ and\ \citenamefont {S\'a~de
  Melo}(2011)}]{SadeMeloC11}%
  \BibitemOpen
  \bibfield  {author} {\bibinfo {author} {\bibfnamefont {K.}~\bibnamefont
  {Seo}}\ and\ \bibinfo {author} {\bibfnamefont {C.~A.~R.}\ \bibnamefont
  {S\'a~de Melo}},\ }\href@noop {} { (\bibinfo {year} {2011})},\ \bibinfo
  {note} {arXiv:1105.4365}\BibitemShut {NoStop}%
\bibitem [{\citenamefont {Sommer}\ \emph {et~al.}(2011)\citenamefont {Sommer},
  \citenamefont {Ku}, \citenamefont {Roati},\ and\ \citenamefont
  {Zwierlein}}]{ZwierleinNature11}%
  \BibitemOpen
  \bibfield  {author} {\bibinfo {author} {\bibfnamefont {A.}~\bibnamefont
  {Sommer}}, \bibinfo {author} {\bibfnamefont {M.}~\bibnamefont {Ku}}, \bibinfo
  {author} {\bibfnamefont {G.}~\bibnamefont {Roati}}, \ and\ \bibinfo {author}
  {\bibfnamefont {M.~W.}\ \bibnamefont {Zwierlein}},\ }\href@noop {} {\bibfield
   {journal} {\bibinfo  {journal} {Nature},\ }\textbf {\bibinfo {volume}
  {472}},\ \bibinfo {pages} {201} (\bibinfo {year} {2011})}\BibitemShut
  {NoStop}%
\bibitem [{\citenamefont {Ku}\ \emph {et~al.}(2012)\citenamefont {Ku},
  \citenamefont {Sommer}, \citenamefont {Cheuk},\ and\ \citenamefont
  {Zwierlein}}]{ZwierleinScience12}%
  \BibitemOpen
  \bibfield  {author} {\bibinfo {author} {\bibfnamefont {M.~J.~H.}\
  \bibnamefont {Ku}}, \bibinfo {author} {\bibfnamefont {A.~T.}\ \bibnamefont
  {Sommer}}, \bibinfo {author} {\bibfnamefont {L.~W.}\ \bibnamefont {Cheuk}}, \
  and\ \bibinfo {author} {\bibfnamefont {M.~W.}\ \bibnamefont {Zwierlein}},\
  }\href@noop {} {\bibfield  {journal} {\bibinfo  {journal} {Science},\
  }\textbf {\bibinfo {volume} {335}},\ \bibinfo {pages} {563} (\bibinfo {year}
  {2012})}\BibitemShut {NoStop}%
\bibitem [{\citenamefont {Mahan}(2000)}]{Mahanbook}%
  \BibitemOpen
  \bibfield  {author} {\bibinfo {author} {\bibfnamefont {G.~D.}\ \bibnamefont
  {Mahan}},\ }\href@noop {} {\emph {\bibinfo {title} {Many-Particle
  Physics}}},\ \bibinfo {edition} {3rd}\ ed.\ (\bibinfo  {publisher} {Kluwer
  academic/Plenum publishers},\ \bibinfo {address} {New York},\ \bibinfo {year}
  {2000})\BibitemShut {NoStop}%
\bibitem [{\citenamefont {Chen}\ \emph {et~al.}(2009)\citenamefont {Chen},
  \citenamefont {He}, \citenamefont {Chien},\ and\ \citenamefont
  {Levin}}]{OurRoPP}%
  \BibitemOpen
  \bibfield  {author} {\bibinfo {author} {\bibfnamefont {Q.~J.}\ \bibnamefont
  {Chen}}, \bibinfo {author} {\bibfnamefont {Y.}~\bibnamefont {He}}, \bibinfo
  {author} {\bibfnamefont {C.~C.}\ \bibnamefont {Chien}}, \ and\ \bibinfo
  {author} {\bibfnamefont {K.}~\bibnamefont {Levin}},\ }\href@noop {}
  {\bibfield  {journal} {\bibinfo  {journal} {Rep. Prog. Phys.},\ }\textbf
  {\bibinfo {volume} {72}},\ \bibinfo {pages} {122501} (\bibinfo {year}
  {2009})}\BibitemShut {NoStop}%
\bibitem [{\citenamefont {Levin}\ \emph {et~al.}(2010)\citenamefont {Levin},
  \citenamefont {Chen}, \citenamefont {Chien},\ and\ \citenamefont
  {He}}]{OurAnnPhys}%
  \BibitemOpen
  \bibfield  {author} {\bibinfo {author} {\bibfnamefont {K.}~\bibnamefont
  {Levin}}, \bibinfo {author} {\bibfnamefont {Q.~J.}\ \bibnamefont {Chen}},
  \bibinfo {author} {\bibfnamefont {C.~C.}\ \bibnamefont {Chien}}, \ and\
  \bibinfo {author} {\bibfnamefont {Y.}~\bibnamefont {He}},\ }\href@noop {}
  {\bibfield  {journal} {\bibinfo  {journal} {Ann. Phys.},\ }\textbf {\bibinfo
  {volume} {325}},\ \bibinfo {pages} {233} (\bibinfo {year}
  {2010})}\BibitemShut {NoStop}%
\bibitem [{\citenamefont {Chien}\ \emph {et~al.}(2012)\citenamefont {Chien},
  \citenamefont {Guo},\ and\ \citenamefont {Levin}}]{OurComment}%
  \BibitemOpen
  \bibfield  {author} {\bibinfo {author} {\bibfnamefont {C.~C.}\ \bibnamefont
  {Chien}}, \bibinfo {author} {\bibfnamefont {H.}~\bibnamefont {Guo}}, \ and\
  \bibinfo {author} {\bibfnamefont {K.}~\bibnamefont {Levin}},\ }\href@noop {}
  {\bibfield  {journal} {\bibinfo  {journal} {Phys. Rev. Lett.},\ }\textbf
  {\bibinfo {volume} {109}},\ \bibinfo {pages} {118901} (\bibinfo {year}
  {2012})}\BibitemShut {NoStop}%
\bibitem [{\citenamefont {Guo}\ \emph {et~al.}(2013){\natexlab{c}}\citenamefont
  {Guo}, \citenamefont {He}, \citenamefont {Chien},\ and\ \citenamefont
  {Levin}}]{OurC13}%
  \BibitemOpen
  \bibfield  {author} {\bibinfo {author} {\bibfnamefont {H.}~\bibnamefont
  {Guo}}, \bibinfo {author} {\bibfnamefont {Y.}~\bibnamefont {He}}, \bibinfo
  {author} {\bibfnamefont {C.~C.}\ \bibnamefont {Chien}}, \ and\ \bibinfo
  {author} {\bibfnamefont {K.}~\bibnamefont {Levin}},\ }\href@noop {} {
  (\bibinfo {year} {2013}{\natexlab{c}})},\ \bibinfo {note}
  {arXiv:1306.3189}\BibitemShut {NoStop}%
\bibitem [{\citenamefont {Kadanoff}\ and\ \citenamefont
  {Martin}(1961)}]{Kadanoff61}%
  \BibitemOpen
  \bibfield  {author} {\bibinfo {author} {\bibfnamefont {L.~P.}\ \bibnamefont
  {Kadanoff}}\ and\ \bibinfo {author} {\bibfnamefont {P.~C.}\ \bibnamefont
  {Martin}},\ }\href@noop {} {\bibfield  {journal} {\bibinfo  {journal} {Phys.
  Rev.},\ }\textbf {\bibinfo {volume} {124}},\ \bibinfo {pages} {670} (\bibinfo
  {year} {1961})}\BibitemShut {NoStop}%
\bibitem [{\citenamefont {Kulik}\ \emph {et~al.}(1981)\citenamefont {Kulik},
  \citenamefont {Entin-Wohlman},\ and\ \citenamefont {Orbach}}]{KulikJLTP81}%
  \BibitemOpen
  \bibfield  {author} {\bibinfo {author} {\bibfnamefont {I.~O.}\ \bibnamefont
  {Kulik}}, \bibinfo {author} {\bibfnamefont {O.}~\bibnamefont
  {Entin-Wohlman}}, \ and\ \bibinfo {author} {\bibfnamefont {R.}~\bibnamefont
  {Orbach}},\ }\href@noop {} {\bibfield  {journal} {\bibinfo  {journal} {J. Low
  Temp. Phys.},\ }\textbf {\bibinfo {volume} {43}},\ \bibinfo {pages} {591}
  (\bibinfo {year} {1981})}\BibitemShut {NoStop}%
\bibitem [{\citenamefont {Zha}\ \emph {et~al.}(1995)\citenamefont {Zha},
  \citenamefont {Levin},\ and\ \citenamefont {Liu}}]{ZhaPRB95}%
  \BibitemOpen
  \bibfield  {author} {\bibinfo {author} {\bibfnamefont {Y.~Y.}\ \bibnamefont
  {Zha}}, \bibinfo {author} {\bibfnamefont {K.}~\bibnamefont {Levin}}, \ and\
  \bibinfo {author} {\bibfnamefont {D.~Z.}\ \bibnamefont {Liu}},\ }\href@noop
  {} {\bibfield  {journal} {\bibinfo  {journal} {Phys. Rev. B},\ }\textbf
  {\bibinfo {volume} {51}},\ \bibinfo {pages} {6602} (\bibinfo {year}
  {1995})}\BibitemShut {NoStop}%
\bibitem [{\citenamefont {Arseev}\ \emph {et~al.}(2006)\citenamefont {Arseev},
  \citenamefont {Loiko},\ and\ \citenamefont {Fedorov}}]{Arseev}%
  \BibitemOpen
  \bibfield  {author} {\bibinfo {author} {\bibfnamefont {P.~I.}\ \bibnamefont
  {Arseev}}, \bibinfo {author} {\bibfnamefont {S.~O.}\ \bibnamefont {Loiko}}, \
  and\ \bibinfo {author} {\bibfnamefont {N.~K.}\ \bibnamefont {Fedorov}},\
  }\href@noop {} {\bibfield  {journal} {\bibinfo  {journal} {Phys. Usp.},\
  }\textbf {\bibinfo {volume} {49}},\ \bibinfo {pages} {1} (\bibinfo {year}
  {2006})}\BibitemShut {NoStop}%
\bibitem [{\citenamefont {Guo}\ \emph {et~al.}(2012)\citenamefont {Guo},
  \citenamefont {Chien},\ and\ \citenamefont {He}}]{OurPRD}%
  \BibitemOpen
  \bibfield  {author} {\bibinfo {author} {\bibfnamefont {H.}~\bibnamefont
  {Guo}}, \bibinfo {author} {\bibfnamefont {C.~C.}\ \bibnamefont {Chien}}, \
  and\ \bibinfo {author} {\bibfnamefont {Y.}~\bibnamefont {He}},\ }\href@noop
  {} {\bibfield  {journal} {\bibinfo  {journal} {Phys. Rev. D},\ }\textbf
  {\bibinfo {volume} {85}},\ \bibinfo {pages} {074025} (\bibinfo {year}
  {2012})}\BibitemShut {NoStop}%
\bibitem [{\citenamefont {Nambu}(1960)}]{Nambu60}%
  \BibitemOpen
  \bibfield  {author} {\bibinfo {author} {\bibfnamefont {Y.}~\bibnamefont
  {Nambu}},\ }\href@noop {} {\bibfield  {journal} {\bibinfo  {journal} {Phys.
  Rev.},\ }\textbf {\bibinfo {volume} {117}},\ \bibinfo {pages} {648} (\bibinfo
  {year} {1960})}\BibitemShut {NoStop}%
\bibitem [{\citenamefont {Schrieffer}(1964)}]{Schrieffer_book}%
  \BibitemOpen
  \bibfield  {author} {\bibinfo {author} {\bibfnamefont {J.~R.}\ \bibnamefont
  {Schrieffer}},\ }\href@noop {} {\emph {\bibinfo {title} {Theory of
  superconductivity}}}\ (\bibinfo  {publisher} {Benjamin},\ \bibinfo {address}
  {New York},\ \bibinfo {year} {1964})\BibitemShut {NoStop}%
\bibitem [{\citenamefont {Leggett}(1980)}]{Leggett}%
  \BibitemOpen
  \bibfield  {author} {\bibinfo {author} {\bibfnamefont {A.~J.}\ \bibnamefont
  {Leggett}},\ }in\ \href@noop {} {\emph {\bibinfo {booktitle} {Modern Trends
  in the Theory of Condensed Matter}}}\ (\bibinfo  {publisher}
  {Springer-Verlag},\ \bibinfo {address} {Berlin},\ \bibinfo {year} {1980})\
  pp.\ \bibinfo {pages} {13--27}\BibitemShut {NoStop}%
\bibitem [{\citenamefont {Mihaila}\ \emph {et~al.}(2011)\citenamefont
  {Mihaila}, \citenamefont {Dawson}, \citenamefont {Cooper}, \citenamefont
  {Chien},\ and\ \citenamefont {Timmermans}}]{largeNcrossover}%
  \BibitemOpen
  \bibfield  {author} {\bibinfo {author} {\bibfnamefont {B.}~\bibnamefont
  {Mihaila}}, \bibinfo {author} {\bibfnamefont {J.~F.}\ \bibnamefont {Dawson}},
  \bibinfo {author} {\bibfnamefont {F.}~\bibnamefont {Cooper}}, \bibinfo
  {author} {\bibfnamefont {C.~C.}\ \bibnamefont {Chien}}, \ and\ \bibinfo
  {author} {\bibfnamefont {E.}~\bibnamefont {Timmermans}},\ }\href@noop {}
  {\bibfield  {journal} {\bibinfo  {journal} {Phys. Rev. A},\ }\textbf
  {\bibinfo {volume} {83}},\ \bibinfo {pages} {053637} (\bibinfo {year}
  {2011})}\BibitemShut {NoStop}%
\bibitem [{\citenamefont {Combescot}\ \emph {et~al.}(2006)\citenamefont
  {Combescot}, \citenamefont {Kagan},\ and\ \citenamefont
  {Stringari}}]{Stringari06}%
  \BibitemOpen
  \bibfield  {author} {\bibinfo {author} {\bibfnamefont {R.}~\bibnamefont
  {Combescot}}, \bibinfo {author} {\bibfnamefont {M.~Y.}\ \bibnamefont
  {Kagan}}, \ and\ \bibinfo {author} {\bibfnamefont {S.}~\bibnamefont
  {Stringari}},\ }\href@noop {} {\bibfield  {journal} {\bibinfo  {journal}
  {Phys. Rev. A},\ }\textbf {\bibinfo {volume} {74}},\ \bibinfo {pages}
  {042717} (\bibinfo {year} {2006})}\BibitemShut {NoStop}%
\bibitem [{\citenamefont {Kubo}\ \emph {et~al.}(2004)\citenamefont {Kubo},
  \citenamefont {Toda},\ and\ \citenamefont {Hashitsume}}]{Kubo_book}%
  \BibitemOpen
  \bibfield  {author} {\bibinfo {author} {\bibfnamefont {R.}~\bibnamefont
  {Kubo}}, \bibinfo {author} {\bibfnamefont {M.}~\bibnamefont {Toda}}, \ and\
  \bibinfo {author} {\bibfnamefont {N.}~\bibnamefont {Hashitsume}},\
  }\href@noop {} {\emph {\bibinfo {title} {Statistical Physics II:
  Nonequilibrium Statistical Mechanics}}},\ \bibinfo {edition} {2nd}\ ed.\
  (\bibinfo  {publisher} {Springer-Verlag},\ \bibinfo {year}
  {2004})\BibitemShut {NoStop}%
\bibitem [{\citenamefont {Yoshimi}\ \emph {et~al.}(2009)\citenamefont
  {Yoshimi}, \citenamefont {Kato},\ and\ \citenamefont
  {Maebashi}}]{Maebashi09}%
  \BibitemOpen
  \bibfield  {author} {\bibinfo {author} {\bibfnamefont {K.}~\bibnamefont
  {Yoshimi}}, \bibinfo {author} {\bibfnamefont {T.}~\bibnamefont {Kato}}, \
  and\ \bibinfo {author} {\bibfnamefont {H.}~\bibnamefont {Maebashi}},\
  }\href@noop {} {\bibfield  {journal} {\bibinfo  {journal} {J. Phys. Soc.
  Jpn.},\ }\textbf {\bibinfo {volume} {78}},\ \bibinfo {pages} {104002}
  (\bibinfo {year} {2009})}\BibitemShut {NoStop}%
\bibitem [{\citenamefont {Pao}\ \emph {et~al.}(2006)\citenamefont {Pao},
  \citenamefont {Wu},\ and\ \citenamefont {Yip}}]{Pao}%
  \BibitemOpen
  \bibfield  {author} {\bibinfo {author} {\bibfnamefont {C.~H.}\ \bibnamefont
  {Pao}}, \bibinfo {author} {\bibfnamefont {S.~T.}\ \bibnamefont {Wu}}, \ and\
  \bibinfo {author} {\bibfnamefont {S.~K.}\ \bibnamefont {Yip}},\ }\href@noop
  {} {\bibfield  {journal} {\bibinfo  {journal} {Phys. Rev. B},\ }\textbf
  {\bibinfo {volume} {73}},\ \bibinfo {pages} {132506} (\bibinfo {year}
  {2006})}\BibitemShut {NoStop}%
\bibitem [{\citenamefont {Chen}\ \emph {et~al.}(2006)\citenamefont {Chen},
  \citenamefont {He}, \citenamefont {Chien},\ and\ \citenamefont
  {Levin}}]{Stability}%
  \BibitemOpen
  \bibfield  {author} {\bibinfo {author} {\bibfnamefont {Q.~J.}\ \bibnamefont
  {Chen}}, \bibinfo {author} {\bibfnamefont {Y.}~\bibnamefont {He}}, \bibinfo
  {author} {\bibfnamefont {C.-C.}\ \bibnamefont {Chien}}, \ and\ \bibinfo
  {author} {\bibfnamefont {K.}~\bibnamefont {Levin}},\ }\href@noop {}
  {\bibfield  {journal} {\bibinfo  {journal} {Phys. Rev. A},\ }\textbf
  {\bibinfo {volume} {74}},\ \bibinfo {pages} {063603} (\bibinfo {year}
  {2006})}\BibitemShut {NoStop}%
\bibitem [{\citenamefont {He}\ \emph {et~al.}(2006)\citenamefont {He},
  \citenamefont {Jin},\ and\ \citenamefont {Zhuang}}]{HePRBS06}%
  \BibitemOpen
  \bibfield  {author} {\bibinfo {author} {\bibfnamefont {L.~Y.}\ \bibnamefont
  {He}}, \bibinfo {author} {\bibfnamefont {M.}~\bibnamefont {Jin}}, \ and\
  \bibinfo {author} {\bibfnamefont {P.~F.}\ \bibnamefont {Zhuang}},\
  }\href@noop {} {\bibfield  {journal} {\bibinfo  {journal} {Phys. Rev. B},\
  }\textbf {\bibinfo {volume} {74}},\ \bibinfo {pages} {214516} (\bibinfo
  {year} {2006})}\BibitemShut {NoStop}%
\end{thebibliography}
%merlin.mbs 2010-03-15 4.21a (PWD, AO, DPC)

\end{document}